\documentclass[12pt]{article}
\usepackage{epsf}
\usepackage[dvips]{graphicx,psfrag}

\renewcommand{\thefootnote}{\#\arabic{footnote}}
\renewcommand{\theequation}{\thesection.\arabic{equation}}

\newcommand{\bear}{\begin{array}}  \newcommand{\eear}{\end{array}}
\newcommand{\bea}{\begin{eqnarray}}  \newcommand{\eea}{\end{eqnarray}}
\newcommand{\beq}{\begin{equation}}  \newcommand{\eeq}{\end{equation}}
\newcommand{\bef}{\begin{figure}}  \newcommand{\eef}{\end{figure}}
\newcommand{\bec}{\begin{center}}  \newcommand{\eec}{\end{center}}
\newcommand{\non}{\nonumber}

\newcommand{\gtrsim}{ \mathop{}_{\textstyle \sim}^{\textstyle >} }
\newcommand{\lesssim}{ \mathop{}_{\textstyle \sim}^{\textstyle <} }

\begin{document}

\renewcommand{\thefootnote}{\fnsymbol{footnote}}
\setcounter{footnote}{0}
\begin{titlepage}

\def\thefootnote{\fnsymbol{footnote}}

\begin{center}

\hfill DESY-03-097\\
\hfill RESCEU-92/03\\
\hfill TU-692\\
\hfill hep-ph/0308174\\
\hfill August, 2003\\

\vskip .5in

{\Large \bf

Curvatons in Supersymmetric Models

}

\vskip .45in

{\large
 Koichi Hamaguchi$^{(a)}$,
 Masahiro Kawasaki$^{(b)}$,\\
 Takeo Moroi$^{(c)}$ and
 Fuminobu Takahashi$^{(b)}$
}

\vskip .3in

{\em $^{(a)}$Deutsches Elektronen-Synchrotron DESY, D-22603, Hamburg,
Germany
}

\vskip .1in

{\em $^{(b)}$Research Center for the Early Universe, School of Science,
University of Tokyo\\
Tokyo 113-0033, Japan}
 
\vskip .1in

{\em $^{(c)}$ Department of Physics, Tohoku University,  Sendai
980-8578, Japan}

\end{center}

\vskip .4in

\begin{abstract}

We study the curvaton scenario in supersymmetric framework paying
particular attention to the fact that scalar fields are inevitably
complex in supersymmetric theories.  If there are more than one scalar
fields associated with the curvaton mechanism, isocurvature (entropy)
fluctuations between those fields in general arise, which may
significantly affect the properties of the cosmic density
fluctuations.  We examine several candidates for the curvaton in the
supersymmetric framework, such as moduli fields, Affleck-Dine field,
$F$- and $D$-flat directions, and right-handed sneutrino.  We estimate
how the isocurvature fluctuations generated in each case affect the
cosmic microwave background angular power spectrum. With the use of
the recent observational result of the WMAP, stringent constraints on
the models are derived and, in particular, it is seen that large
fraction of the parameter space is excluded if the Affleck-Dine field
plays the role of the curvaton field.  Natural and well-motivated
candidates of the curvaton are also listed.

\end{abstract}

%\rem{Comment on the exclusion of the AD case: some figure or ???}

%\rem{Comment on the photon from inflaton}

\end{titlepage}

\renewcommand{\thepage}{\arabic{page}}
\setcounter{page}{1}
\renewcommand{\thefootnote}{\#\arabic{footnote}}
\setcounter{footnote}{0}

\renewcommand{\theequation}{\thesection.\arabic{equation}}

\section{Introduction}
\label{sec:introduction}
\setcounter{equation}{0}

Study of the origin of the cosmic density fluctuations is a very
important subject in cosmology.  In the conventional scenarios,
inflation is assumed as a mechanism to provide the source of the
cosmic density fluctuations.  In the inflationary scenarios, the
universe is assumed to experience the epoch of de Sitter expansion in
the early stage.  During the de Sitter expansion, physical scale
expands faster than the horizon scale so the homogeneity of the
universe is realized at the classical level.  At the quantum level,
however, scalar field responsible for the inflation, called
``inflaton,'' acquires quantum fluctuation, which becomes the origin
of the cosmic density fluctuations.  If inflation provides the source
of the cosmic density fluctuations, energy scale of the inflation is
related to the amplitude of the cosmic density fluctuations.
Importantly, observations of the cosmic microwave background (CMB)
anisotropy give us important informations about the inflationary
models.  In particular, in the simplest scenarios, inflation models
should reproduce the currently observed size of the CMB
anisotropy of $\Delta T/T\sim O(10^{-5})$.  Furthermore, spectral
index for the scalar-mode perturbations should be close to $1$ so that
the shape of the CMB angular power spectrum becomes consistent with
observations.  After the very precise measurement of the CMB angular
power spectrum by the Wilkinson Microwave Anisotropy Probe (WMAP)
\cite{wmap}, thus we obtain stringent constraints on the inflation
models.

Recently, a new mechanism is proposed where a late-decaying scalar
condensation becomes the dominant source of the cosmic density
fluctuations \cite{NPB626-395,PLB524-5,MorTak,PRD67-023503}.  In this
scenario, a scalar field, called ``curvaton,'' other than the
inflaton, acquires primordial fluctuations in the early universe.
Although the energy density of the curvaton is subdominant in the
early stage, it eventually dominates the universe.  Consequently, the
isocurvature (entropy) fluctuations originally stored in the curvaton
sector become the adiabatic ones \cite{iso2adi} which become the
dominant source of the density fluctuations of the universe.  Then,
when the curvaton decays, the universe is reheated and the fluctuation
of the curvaton produces the adiabatic density fluctuations.

The curvaton mechanism has important implications to particle
cosmology.  If the curvaton mechanism is implemented in inflation
models, the size and the scale dependence of the cosmological density
fluctuations become different from the conventional result.  In
particular, the curvaton mechanism provides a natural scenario of
generating (almost) scale-invariant cosmic density fluctuations which
is strongly suggested by observations.  As a result, we can relax the
observational constraints on the inflation models
\cite{PLB524-5,MorTak,PRD67-023503}.

Various particle-physics candidates of the curvaton field have been
discussed so far.  (For the recent discussions on the curvaton
scenario, see Ref.\ \cite{RecentCurvaton}.)  Among them, it is often
the case that the curvaton mechanism is considered in the framework of
supersymmetry, since the supersymmetry can protect the flatness of the
curvaton potential which is required for a successful curvaton
scenario.  A crucial point in supersymmetric theories is that scalar
fields are inevitably complex and hence they have two (independent)
degrees of freedom.  Thus, if the curvaton mechanism is implemented in
supersymmetric models, effects of two fields should be carefully taken
into account.  In particular, if there are two scalar fields,
isocurvature fluctuations between them are in general induced which
may significantly affect the behavior of the cosmic density
fluctuations.

Thus, in this paper, we study implications of the curvaton scenarios
in the framework of supersymmetric models.  We pay a special attention
to the fact that, in the supersymmetric framework, curvaton mechanism
requires at least two (real) scalar fields.  The CMB anisotropy in
such a framework is studied in detail and we investigate how the CMB power
spectrum behaves.

The organization of the rest of this paper is as follows.  In Section
\ref{sec:framework}, we first present the framework of the curvaton
mechanism.  Several basic issues concerning the cosmic density
fluctuations are discussed in Section \ref{sec:perturbations}.  Then,
cases where the cosmological moduli fields, Affleck-Dine field, $F$-
and $D$-flat directions, and the right-handed sneutrino play the role
of the curvaton are discussed in Sections \ref{sec:moduli},
\ref{sec:ad}, \ref{sec:flatdir} and \ref{sec:snu}, respectively.
Section \ref{sec:conclusion} is devoted for the conclusions and
discussion.

\section{Framework}
\label{sec:framework}
\setcounter{equation}{0}

We first present the framework.  In the scenario we consider there are
two classes of scalar fields which play important roles.  The first
one is the inflaton field which causes  inflation.\footnote
{Alternative models such as the pre-big-bang \cite{pbb,pbb2} and the
ekpyrotic \cite{ekpyrotic} scenarios were proposed.  However, they
predict unwanted spectrum for the density fluctuations~\cite{pbb2,ekpy2}.  
Thus, we assume inflation in the following discussion.}
In the conventional scenarios, the inflaton field is assumed to be
responsible for the source of the cosmic density fluctuations but here
this is not the case.  The second one is the curvaton which becomes
the origin of the cosmic density fluctuations.  Here, we consider
general cases where there exist more than one curvaton fields.
Hereafter curvaton is denoted as $\phi$ (with relevant subscripts as
mentioned below).

In this paper, we consider the case where the universe starts with the
inflationary epoch.  During inflation, the universe exponentially
expands and the horizon and flatness problems are solved.  In
addition, the curvaton fields are assumed to have non-vanishing
amplitude during inflation.  Assuming that the masses of the
curvaton fields are much smaller than the expansion rate of the
universe during inflation, the energy density of the curvatons are
minor component in the very early universe.

After inflation, the universe is reheated by the decay of the inflaton
field and radiation dominated universe is realized.  (We call this
epoch as ``RD1'' epoch.)  At the time of the reheating, the curvaton
fields are minor components and their energy densities are negligibly
small compared to that of the radiation.  As the universe expands, the
expansion rate decreases and the curvatons start to oscillate when the
Hubble parameter becomes comparable to the masses of them. Then, the
(averaged) amplitudes decrease.

In discussing the evolution of the curvatons, it is convenient to
decompose the complex field into real ones.  One convenient convention
is to use the mass-eigenstate basis.  When the amplitudes of the
scalar fields become small enough, the potential of $\phi$ can be
approximated by the parabolic one as $V\simeq
m^2|\hat{\phi}|^2+m'^2(\hat{\phi}^2+{\rm h.c.})$.  (Notice that $m'^2$
can be always chosen to be real using the phase rotation of the
$\hat{\phi}$ field.)  Then, the real and imaginary parts of
$\hat{\phi}$ become the mass eigenstates.  Expanding $\hat{\phi}$ as
\begin{eqnarray}
    \hat{\phi} =
    \frac{1}{\sqrt{2}} \left( \phi_1 + i \phi_2 \right),
    \label{phihat->phi12}
\end{eqnarray}
the scalar potential is described by the quadratic form around its
minimum as
\begin{eqnarray}
    V = \frac{1}{2} \sum_i m_{\phi_i}^2 \phi_i^2,
    \label{V=m^2phi^2}
\end{eqnarray}
where $\phi_i$ denotes $i$-th mass eigenstate with $i=1,2$.  Here and
hereafter, the ``hat'' is for complex scalar fields while the scalars
without the ``hat'' are understood to be real.  In addition, the
indices $i$, $j$, $\cdots$ are for specifying the mass eigenstates.
In general, this basis should be distinguished from the ones expanding
the fluctuation during the inflation.  Hereafter, we assume that the
masses of the $\phi_i$ fields are of the same order of magnitude.

Although the mass-eigenstate basis is useful in particular in
discussing the behavior of the curvaton fields when the amplitudes
become small, it is sometimes more convenient to use the
``polar-coordinate basis'' where $\hat{\phi}$ is decomposed as
\begin{eqnarray}
\label{eq:radialrep}
   \hat{\phi} = \frac{1}{\sqrt{2}} 
   \left( \phi_{\rm init} + \phi_r\right) 
   e^{i (\theta_{\rm init} + \phi_\theta / \phi_{\rm init})}.
\end{eqnarray}
Here, $\phi_{\rm init}/\sqrt{2}$ and $\theta_{\rm init}$ correspond to
the initial amplitude and phase of the complex scalar field
$\hat{\phi}$, respectively. In addition, $\phi_r$ and $\phi_\theta$
are real scalar fields.

Adopting the quadratic potential, the (averaged) pressure of each
curvaton vanishes once the curvaton field starts to oscillate:
\begin{eqnarray}
   p_{\phi_i} = \left\langle 
  \frac{1}{2}\dot{\phi}_i^2 - \frac{1}{2}m_{\phi_i}^2 \phi_i^2 
   \right\rangle_t =0,
\end{eqnarray}
where the ``dot'' denotes the derivative with respect to time $t$.
This means that, at this epoch, the energy density of the curvatons
behave as that of non-relativistic matter, namely
$\rho_{\phi_i}\propto a^{-3}$ with $a$ being the scale factor, while
the energy density of the radiation drops faster, $\rho_{\gamma}
\propto a^{-4}$. Thus the curvaton fields can dominate the universe,
which is one of the indispensable conditions for the curvaton scenario
to work.  (We call this epoch as ``$\phi$ dominated'' or ``$\phi$D''
epoch.)  Then, the curvaton fields decay and the universe is reheated
again.  Consequently, the universe is dominated by the radiations
generated from the decay products of the curvatons.  (We call the
second radiation dominated epoch as ``RD2'' epoch.)

\section{Density Fluctuations and CMB Power Spectrum}
\label{sec:perturbations}
\setcounter{equation}{0}

During the inflationary epoch, all the scalar-field amplitudes acquire
quantum fluctuations.  If one calculates the two point correlation
function of a scalar field $\varphi$ with mass $m_\varphi$:
\begin{eqnarray}
    \langle 0 | \delta\varphi (t, \vec{x}) \delta\varphi (t, \vec{y}) 
    | 0 \rangle_{\rm inf} =
    \int \frac{d k}{k}
    |\delta\varphi (t, \vec{k})|^2
    e^{i \vec{k} (\vec{x}-\vec{y})},
\end{eqnarray}
where $\vec{x}$ and $\vec{y}$ are comoving coordinate, then the
Fourier amplitude $\delta\varphi (t, \vec{k})$ is given by
\begin{eqnarray}
    \delta\varphi (t, \vec{k}) = 
    \left[ \frac{H_{\rm inf}}{2\pi} \right]_{k=aH_{\rm inf}}
    \times \left\{ 
    \begin{array} {ll}
    (k/2aH_{\rm inf})^{m_\varphi^2/3H_{\rm inf}^2} 
    &~~~:~~~ 
    m_{\varphi} \ll H_{\rm inf}\\
    (k/aH_{\rm inf})^{3/2}
    \sqrt{H_{\rm inf}/m_\varphi}
    &~~~:~~~ 
    m_{\varphi} \gg H_{\rm inf} 
    \end{array}\right.  ,
    \label{dvarphi}
\end{eqnarray}
where $H_{\rm inf}$ is the expansion rate of the universe during the
inflation and $k$ is the comoving momentum.  For the fluctuations we
are interested in, the wavelengths are much longer than the horizon
scale during inflation.

As one can see from Eq.\ (\ref{dvarphi}), for the superhorizon mode
({\it i.e.}, for $k\ll aH_{\rm inf}$), fluctuations of the
scalar-field amplitude are suppressed when the scalar mass is
comparable to or larger than $H_{\rm inf}$.  In inflation models in
the framework of the supergravity, it is often the case that the
effective mass of $O(H_{\rm inf})$ is generated to flat directions.
(Such a scalar mass is called the ``Hubble-induced scalar mass.'')  If
this is the case, the curvaton mechanism does not work.  However, even
in the supergravity, there are several cases where curvaton fields
(and other flat directions) do not acquire Hubble-induced masses.  One
case is the so-called $D$-term inflation \cite{DTInf} where the
``vacuum energy'' during the inflation is provided by a $D$-term
potential.  In this case, flatness of the curvaton potential is not
severely disturbed since the Hubble-induced scalar masses originate
from the $F$-term interaction.\footnote
{After inflation, the inflaton starts oscillation, which induces
$F$-term potential. Thus, the Hubble-induced mass term appears even for
$D$-term inflation, and the curvaton fluctuations decrease as
$a^{-3/2}$ until reheating.  Importantly, the amplitude itself is also
proportional to $a^{-3/2}$ with the parabolic potential.  Thus, the
ratio $\delta\phi/\phi$ becomes a constant of time and the effects of
the Hubble-induced mass term do not change the following discussion.}
Another possibility of maintaining the flatness of the curvaton
potential is the no-scale type inflation.  If the K\"ahler potential
is in the no-scale form, it is known that the Hubble-induced mass is
suppressed \cite{PLB355-71}.  The no-scale type K\"ahler potential can
be naturally realized in the sequestered sector scenario
\cite{NPB557-79} where the inflaton sector lives in the hidden brane
which is geometrically separated from the observable brane where the
standard-model particles live.\footnote
{Such a set-up may be naturally fit into the scenario of the
anomaly-mediated supersymmetry breaking
\cite{NPB557-79,JHEP9812-027,JHEP0004-009}, since the hidden and
observable branes are usually introduced in the anomaly-mediated
supersymmetry breaking models to explain the smallness of the
K\"ahler-induced supersymmetry breaking scalar masses.}
In the conventional inflation scenario, in fact, such a scenario is
problematic since the decay of the inflaton field reheats the hidden
brane not the observable brane.  In the curvaton scenario, however,
this is not a problem since dominant part of the matter in the
universe ({\it i.e.}, radiation, cold dark matter (CDM), baryon
asymmetry, and so on) is generated from the decay of the curvaton
fields.\footnote
{In scenarios with late-time entropy production, in fact, this is a
general property and the inflaton field does not have to decay into
the standard-model particles.  Thus, in this case, the inflaton can be
a hidden-sector field and its interaction with the standard-model
particles can be absent.}

The curvatons also acquire quantum fluctuations during the inflation.
As we mentioned before, the curvaton is expected to be a complex field
in the supersymmetric framework and hence there are two (or more)
independent real scalar fields associated with the curvaton mechanism.
If the flatness of the curvaton potential is not disturbed by the
inflation dynamics, all the curvatons are expected to acquire the
quantum fluctuations as given in Eq.\ (\ref{dvarphi}) with $m_\varphi
\ll H_{\rm inf}$. In the following, to make our discussion clearer, we
consider the case where there are two independent real scalar fields,
corresponding to the real and imaginary parts of the complex scalar
field.  (Our formulae can be easily generalized to the cases with more
scalar fields.)

During inflation, we expand the curvaton $\hat{\phi}$ around its
amplitude $[\hat{\phi}]_{\rm inf}$:
\begin{eqnarray}
   \hat{\phi} = [ \hat{\phi} ]_{\rm inf} + \delta \hat{\phi}.
\end{eqnarray}
In general, there are two (uncorrelated) real scalar fields
parameterizing $\delta\hat{\phi}$.  We denote the fluctuations of such
scalar fields as $\delta\phi_A$.  (Hereafter, the indices $A$, $B$,
$\cdots$ are for specifying the basis expanding the scalar field
during inflation.)

One way of parameterizing $\delta\hat{\phi}$ is to use the mass basis.
When the amplitude of the scalar field is small enough, the potential
can be well approximated by the parabolic one as Eq.\ 
(\ref{V=m^2phi^2}), which enables us to use this basis in order to
parameterize $\delta\hat{\phi}$.  One should note that, if the scalar
potential has interaction terms, primordial fluctuations in $\phi_i$
may generate fluctuations in $\phi_j$ with $i\neq j$.  Of course, if
the potential is completely parabolic, this does not happen.  

Importantly, fluctuations of two independent scalar fields are
uncorrelated.  Thus, in calculating the two point correlation function
of the cosmic temperature fluctuations from which the CMB anisotropy is
obtained, we can regard (Fourier modes of) the fluctuations of the
curvatons as independent statistical variables, leading to
\begin{eqnarray}
\left\langle \delta\phi_A \delta\phi_B \right\rangle
= \left[ \frac{H_{\rm inf}}{2\pi} \right]^2 \delta_{AB}.
\label{<AB>}
\end{eqnarray}
These fluctuations become the origin of the cosmic density
fluctuations.  If $\delta\phi_A$ and $\delta\phi_B$ are independent,
the resultant CMB angular power spectrum is the same for any choice of
the basis.  As we will see below, they may provide both adiabatic and
isocurvature fluctuations and hence a large class of scenarios are
severely constrained from the observations of the CMB angular power
spectrum.

Once the fluctuations of the curvaton amplitudes are generated as in
Eq.\ (\ref{<AB>}), the density and metric perturbations are induced
from $[\delta\phi]_{\rm inf}$.  To parameterize the density
fluctuations, we define the variable\footnote
{We adopt the notation and convention used in Ref.\ \cite{hu_PhD}
unless otherwise mentioned.}
\begin{eqnarray}
   \delta_X \equiv \frac{\delta\rho_X}{\rho_X},
\end{eqnarray}
where the subscript $X$ is to distinguish various components.
Hereafter, $X=\gamma$, $c$, $b$, and $m$ are for the photon (or more
precisely, relativistic matter), CDM, baryon, and total
non-relativistic matter, respectively.  The density fluctuation
$\delta\rho_X$ is defined in the Newtonian gauge where the perturbed
line element is given using the metric perturbations $\Psi$ and $\Phi$
as
\begin{eqnarray}
    ds^2 &=& - (1 + 2\Psi) dt^2
    + a^2 (1 + 2\Phi) \delta_{ij} dx^i dx^j 
    \nonumber \\
     &=& 
    a^2 \left[ - (1 + 2\Psi) d\tau^2 + (1 + 2\Phi) 
    \delta_{ij} dx^i dx^j \right].
    \label{ds^2}
\end{eqnarray}

Evolutions of the metric and density perturbations are governed by the
Einstein and Boltzmann equations.  The Einstein equation provides the
relation between metric and density perturbations:
\begin{eqnarray}
    k^2 \Phi 
    = \frac{1}{2M_*^2} a^2
    \rho_{\rm T} \left[ 
        \delta_{\rm T} + \frac{3 {\cal H}}{k} 
        (1+\omega_{\rm T}) V_{\rm T}
    \right],  
    \label{poisson}
\end{eqnarray}
and the equation relating $\Phi$, $\Psi$, and the anisotropic stress
perturbation of the total matter $\Pi_{\rm T}$:
\begin{eqnarray}
    k^2 (\Psi + \Phi) = - \frac{1}{M_*^2} a^2
    p_{\rm T} \Pi_{\rm T},
    \label{Psi+Phi=0}
\end{eqnarray}
where $M_*\simeq 2.4\times 10^{18}\ {\rm GeV}$ is the reduced Planck
scale.  Here, the subscript ``T'' is for the total matter and the
variables $p_X$ and $V_X$ are the pressure and the velocity
perturbation of the component $X$, respectively.  In addition,
$\omega_{\rm T}\equiv p_{\rm T}/\rho_{\rm T}$ is the equation-of-state
parameter for the total matter, and
\begin{eqnarray}
    {\cal H} \equiv \frac{a'}{a},
\end{eqnarray}
where the ``prime'' denotes the derivative with respect to the
conformal time $\tau$.  

In order to discuss the CMB angular power spectrum associated with the
fluctuations of the curvaton fields, we first evaluate the density
fluctuations at the deep RD2 epoch.  In such an epoch, the mean free
path of the radiation is very short and hence the fluctuation of the
radiation component becomes locally isotropic.  In this case, the
anisotropic stress perturbation vanishes and we obtain
\begin{eqnarray}
    \label{eq:radiation}
    \delta_\gamma' &=& -\frac{4}{3} k V_\gamma - 4 \Phi', 
    \label{dr'} \\
    V_\gamma' &=& \frac{1}{4} k \delta_r + k\Psi.
    \label{Vr'}
\end{eqnarray}
In addition, if the anisotropic stress perturbation vanishes,
$\Phi=-\Psi$ due to Eq.\ (\ref{Psi+Phi=0}).  We use this relation to
eliminate $\Phi$.  For non-relativistic component, the Boltzmann equation
becomes
\begin{eqnarray}
    \delta_m' &=& -k V_m - 3\Phi',
    \label{eq:cdm} \\
    V_m'  &=& - \mathcal{H} V_m + k \Psi.
    \label{eq:cdm_velocity}
\end{eqnarray}

It is convenient to use the mass basis in discussing the density
fluctuations in the RD2 epoch.  Furthermore, it is useful to define
the primordial entropy between $i$-th curvaton and the radiation
generated from the decay of the inflaton field:
\begin{eqnarray}
   S_{i} \equiv 
   \left[ \delta_{\phi_i} - \frac{3}{4} \delta_{\gamma_{\rm inf}}
       \right]_{\rm \phi D},
\end{eqnarray}
where $\delta_{\gamma_{\rm inf}}$ is the density fluctuation of the
radiation generated from the inflaton field.  Notice that, in studying
effects of the primordial fluctuations of the curvaton fields,
$\delta\rho_{\gamma_{\rm inf}}$ vanishes in the RD1 epoch and hence
$S_{i}$ is determined by the initial value of the density fluctuation
of the curvatons.  Since the entropy between $\phi$ and $\gamma_{\rm
inf}$ is conserved for the superhorizon mode, $S_{i}$ becomes the
entropy between the components generated from the decay product of
$\phi$ and those not from $\phi$.  Then, we define the transition
matrix ${\cal T}$ as
\begin{eqnarray}
   S_{i} = {\cal T}_{iA} [ \delta\phi_A ]_{\rm inf}.
   \label{DefofT}
\end{eqnarray}
The transition matrix ${\cal T}$ depends on models and initial
conditions.  If the potentials of the scalar fields are well
approximated by the parabolic one, it is convenient to choose mass
eigenstate as $\delta\phi_A$.  In this case, with relevant unitary
transformation, ${\cal T}$ can be expressed as
\begin{eqnarray}
   {\cal T}_{iA} = 
   [{\rm diag} (2\phi_1^{-1}, 2\phi_2^{-1}, \cdots)]_{iA},
   \label{T-diag}
\end{eqnarray}
or equivalently 
%\cite{MorTak}
%%
\begin{eqnarray}
    S_{i} = \left[ \frac{2\delta\phi_i}{\phi_i} \right]_{\rm inf}.
\end{eqnarray}
In this case, the correlations between the fluctuations of different
mass eigenstates vanish.  For general scalar potential, however,
${\cal T}$ is not guaranteed to be diagonal in this basis and the
correlation function $\langle\delta\phi_i\delta\phi_j\rangle$ may
become non-vanishing for $i\neq j$.  Consequently, the isocurvature
fluctuations for different mass eigenstates can become correlated.

Now we study the evolution of the density and metric fluctuations with
non-vanishing $S_{i}$ with $S_{j}=0$ for $j\neq i$.  (Note that the
effects of individual $S_{i}$ can be treated separately as long as the
linear perturbation theory is valid.)  For this purpose, it is
convenient to expand the fluctuations as functions of $k\tau$ since we
are interested in behaviors of superhorizon modes.  In addition, it is
important to note that, since the curvaton fields are minor components
in the RD1 epoch, the total density fluctuation at that epoch is
negligibly small and hence $[\Psi^{(\delta\phi_A)}]_{\rm RD1}$
vanishes.  (Here and hereafter, the superscript $(\delta\phi_A)$ is
for quantities generated from the primordial fluctuation of $\phi_A$.)
The density fluctuation of the photon generated from the decay product
of $\phi_j$ ($j\neq i$) is obtained by using the fact that the
velocities of the scalar-field condensations are higher order in
$k\tau$ relative to $\Psi$ and $\delta_\phi$.\footnote
{Note that it is possible to define those photons generated from the
decay products of each curvaton and inflaton separately, as far as we
are concerned with the density fluctuation in the long wavelength
limit. The reason for this is that the density fluctuation over the
horizon scale is determined only by the gravitational potential, as
can be seen from Eq.~(\ref{eq:radiation}).  If we take into
consideration the effect of the higher order in $k\tau$, such a
distinction becomes meaningless since those photons constitute a
single component fluid with a common velocity perturbation.}
Thus, from Eqs.\ (\ref{dr'}) and (\ref{Vr'}), we obtain
$\delta_{\gamma_j}^{(\delta\phi_A)}=4\Psi^{(\delta\phi_A)}+O(k^2\tau^2)$
for $j\neq i$, where $\gamma_k$ denotes the photon generated from the
decay products of $\phi_k$.  Also, the density fluctuation of the
photon generated from $\phi_i$ is obtained using the relations
\cite{hu_PhD}
\begin{eqnarray}
   [ \delta_{\rm T} ]_{\rm RD2} =  -2 [ \Psi ]_{\rm RD2},
\end{eqnarray}
and
\begin{eqnarray}
   [ \delta_{\rm T} ]_{\rm RD2} =
   \sum_k f_{\gamma_k} [ \delta_{\gamma_k} ]_{\rm RD2},
   \label{delta_T(RD2)}
\end{eqnarray}
where $f_{\gamma_k}=[\rho_{\gamma_k}/\rho_{\gamma}]_{\rm RD2}$ is the
energy fraction of the radiation generated from the decay products of
$k$-th mass eigenstate.  If $S_{i}$ is the only source of the cosmic
density fluctuations,
$S_{i}=\frac{3}{4}(\delta_{\gamma_i}-\delta_{\gamma_j})$ for $j\neq
i$, and hence we obtain $[\Psi]_{\rm
RD2}^{(\delta\phi_A)}=-\frac{2}{9}f_{\gamma_i}S_{i}^{(\delta\phi_A)}$.

In general, the primordial fluctuation in $\phi_A$ generates
fluctuations in various mass eigenstates and hence, in the RD2 epoch,
the metric perturbation is related to the isocurvature fluctuations as
\begin{eqnarray}
   [ \Psi^{(\delta\phi_A)} ]_{\rm RD2} 
   = -\frac{2}{9} \sum_i f_{\gamma_i} S_{i}^{(\delta\phi_A)},
   \label{Psi(RD2)}
\end{eqnarray}
where $\Psi^{(\delta\phi_A)}$ is the metric perturbation generated from
$[\delta\phi_A]_{\rm inf}$.  In particular, when $i=1$ and $2$ with
$f_{\gamma_1}+f_{\gamma_2}=1$, we can write
\begin{eqnarray}
   [ \Psi ]_{\rm RD2} = 
   -\frac{2}{9} \left[
   \bar{S} + \frac{1}{2} (f_{\gamma_1} - f_{\gamma_2}) 
   S_{12} \right],
   \label{Psi(S)}
\end{eqnarray}
where
\begin{eqnarray}
   \bar{S} \equiv \frac{1}{2} \left( S_{1} + S_{2} \right),~~~
   S_{12} \equiv S_{1} - S_{2}.
   \label{S+/S12}
\end{eqnarray}
Notice that $S_{12}$ becomes the entropy between components generated
from the decay products of $\phi_1$ and $\phi_2$.

In discussing the CMB anisotropy, the CMB angular power spectrum is
usually used:
\begin{eqnarray}
    \left\langle \Delta T(\vec{x}, \vec{\gamma}) 
        \Delta T(\vec{x}, \vec{\gamma}')  \right\rangle_{\vec{x}}
    = \frac{1}{4\pi} 
    \sum_l (2l+1) C_l P_l (\vec{\gamma} \cdot \vec{\gamma}'),
\end{eqnarray}
with $\Delta T (\vec{x}, \vec{\gamma})$ being the temperature
fluctuation of the CMB radiation pointing to the direction
$\vec{\gamma}$ at the position $\vec{x}$ and $P_l$ is the Legendre
polynomial.  In the curvaton scenario, the cosmic density fluctuations
originate from the primordial fluctuations of the curvaton fields.  In
particular, as mentioned before, more than one curvaton fields are
expected in the supersymmetric case. Then, denoting the contribution
of $\delta\phi_A$ as $C_l^{(\delta\phi_A)}$, the CMB angular power
spectrum has the form
\begin{eqnarray}
        \label{eq:clsum}
    C_l & = &  \sum_A 
    [ \Psi^{(\delta\phi_A)} ]^2 
    [ C_l^{(\delta\phi_A)} ]_{\left|\Psi\right| =1}.
\end{eqnarray}
Notice that $C_l^{(\delta\phi_A)}\propto\delta\phi_A^2$.  Since the
fluctuations of different curvatons are uncorrelated as shown in Eq.\ 
(\ref{<AB>}), no term proportional to $\delta\phi_A\delta\phi_B$ with
$A\neq B$ is expected.

The CMB angular power spectrum depends on the isocurvature
fluctuations in the baryon and the CDM as well as the metric
perturbation.  Although the isocurvature fluctuations can be
separately defined for the CDM and for the baryonic component, the
shape of $C_l^{(\delta\phi_A)}$ is, up to normalization, determined by
the ratio
\begin{eqnarray}
   \kappa_m^{(\delta\phi_A)} \equiv 
   \left[ \frac{S_{mr}^{(\delta\phi_A)}}{\Psi^{(\delta\phi_A)}} 
   \right]_{\rm RD2},
\end{eqnarray}
where $S_{mr}$ is given by
\begin{eqnarray}
   S_{mr} = \delta_m - \frac{3}{4}\delta_\gamma = 
   \frac{\Omega_b \delta_b + \Omega_c \delta_c}{\Omega_m} 
   - \frac{3}{4}\delta_\gamma.
\end{eqnarray}
Notice that the isocurvature fluctuations are in general correlated
with the metric perturbations. Thus, the form of the CMB angular power
spectrum for one curvaton case is written in the form
\begin{equation}
    \left[C_{l}^{(\delta\phi_A)} \right]_{\left|\Psi\right| =1}
    = %N^{(\delta\phi_A)} \left( 
    	C_{l}^{\rm (adi)} 
        + 2\kappa_m^{(\delta\phi_A)} C_{l}^{\rm (corr)} 
        + \kappa_m^{(\delta\phi_A)2} C_{l}^{\rm (iso)}.
        % \right).
\end{equation}
Here, $C_{l}^{\rm (adi)}$ and $C_{l}^{\rm (iso)}$ agree with the
angular power spectra for purely adiabatic and isocurvature
fluctuations, respectively, while $C_{l}^{\rm (corr)}$
parameterizes the effects of the correlation. 
Their formal definitions are given as
\bea
C_{l}^{\rm (adi)} &\equiv& \left[C_{l} \right]_{
 	\left|\Psi\right| =1,~ \left| S \right| =0},\non\\
C_{l}^{\rm (iso)} &\equiv& \left[ C_{l} \right]_{
 	\left|\Psi\right| =0,~ \left| S \right| =1},\non\\	
C_{l}^{\rm (corr)} &\equiv& \frac{1}{2} \left(\left[ C_{l} \right]_{\Psi/S = 1,~
 	\left|\Psi\right| = \left| S \right| =1} - C_{l}^{\rm (adi)} -
	C_{l}^{\rm (iso)} \right).
\eea
%%
% In addition, $N^{(\delta\phi_A)}$ is the overall normalization.  
When two (or more) curvatons exist, the above equation is generalized as
\begin{equation}
    C_{l} = \left(\sum_A 
    [ \Psi^{(\delta\phi_A)} ]^2 \right) \left( C_{l}^{\rm (adi)} 
        + 2\kappa_{\rm corr} C_{l}^{\rm (corr)} 
        + \kappa_{\rm iso}^2 C_{l}^{\rm (iso)} \right).
\end{equation}
Combining this equation with the Eq.\ (\ref{eq:clsum}), $\kappa_{\rm
corr}$ and $\kappa_{\rm iso}$ are related to
$\kappa_m^{(\delta\phi_A)}$ as
\begin{eqnarray}
    \kappa_{\rm corr} =
    \frac{
    \sum_A [ \Psi^{(\delta\phi_A)} ]_{\rm RD2}^2 
    \kappa_m^{(\delta\phi_A)}}
    {\sum_A [ \Psi^{(\delta\phi_A)} ]_{\rm RD2}^2},
    ~~~
    \kappa_{\rm iso}^2 = \frac{
    \sum_A [ \Psi^{(\delta\phi_A)} ]_{\rm RD2}^2 
    \kappa_m^{(\delta\phi_A)}{}^2}
    {\sum_A [ \Psi^{(\delta\phi_A)} ]_{\rm RD2}^2}.
    \label{eq:relkappa}
\end{eqnarray}
Note that $\kappa_{\rm iso} \geq \left|\kappa_{\rm corr}\right|$ is always satisfied 
from their definitions.

\begin{figure}
    \centering
    \includegraphics[width=8.5cm]{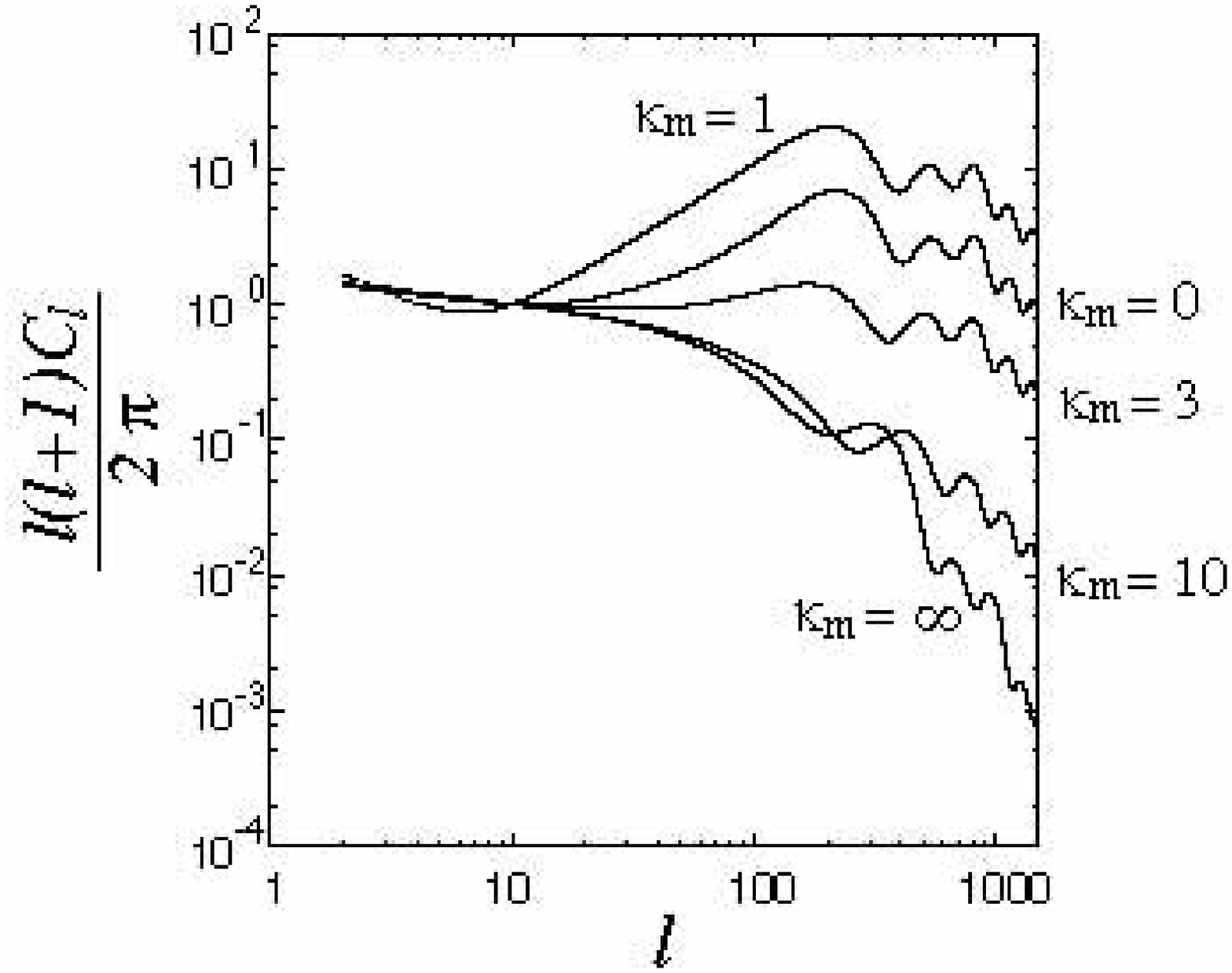}
    \includegraphics[width=8.5cm]{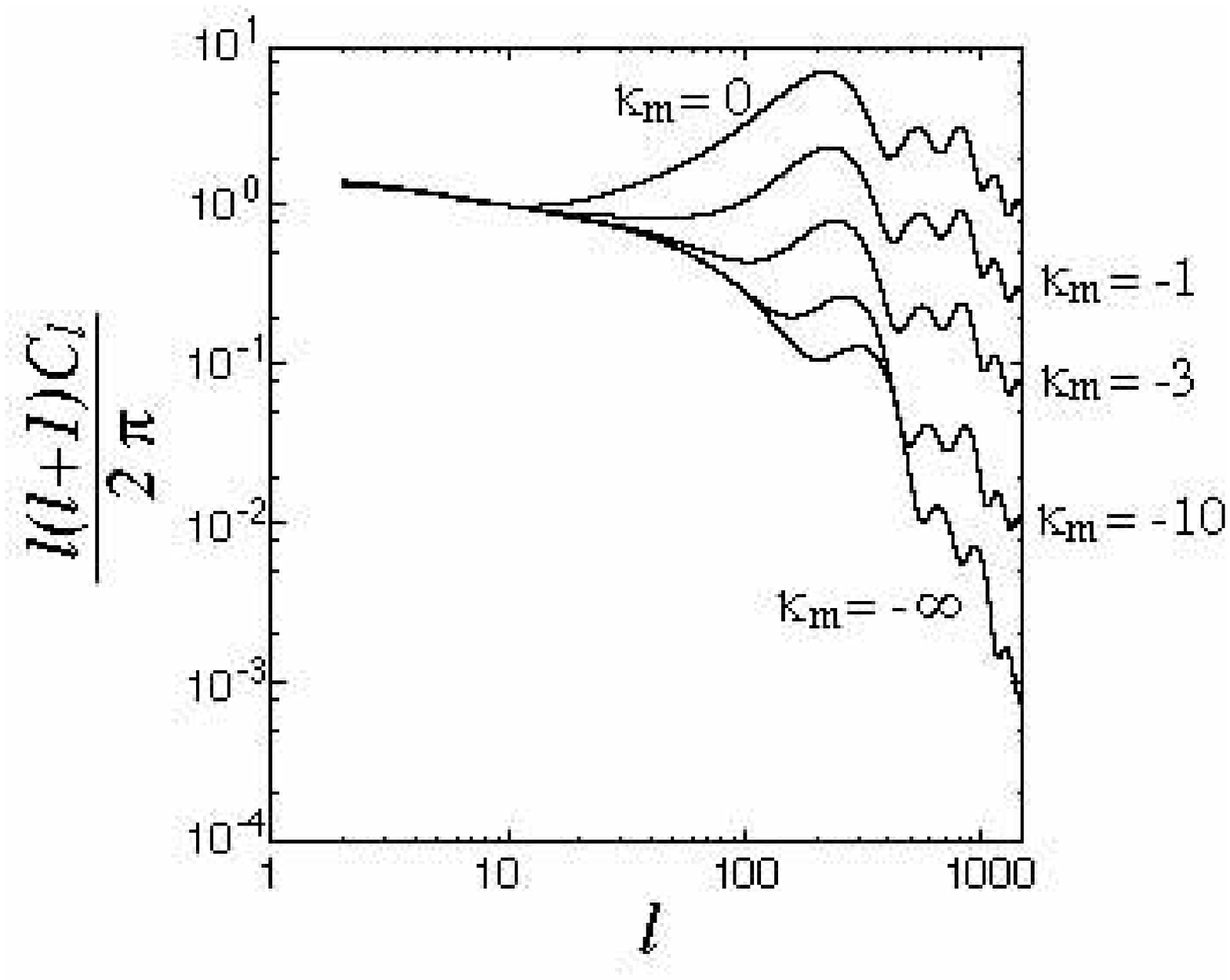}
    \caption{The CMB angular power spectra with correlated 
    mixture of the adiabatic and isocurvature fluctuations.  Here, we
    take $\kappa_m=\infty,~10,~3,~1,~0,~-1,~-3,~-10,~-\infty$. The
    overall normalizations are taken as $[l(l+1)C_l/2\pi]_{l=10}=1$.}
    \label{fig:kappa-dep}
\end{figure}

In Fig.\ \ref{fig:kappa-dep}, we plot the CMB power spectrum $C_l$
for several values of $\kappa_m$ with the relation 
$\kappa_{\rm corr}=\kappa_{\rm iso}=\kappa_m$.
(In our study, we use the cosmological parameters $\Omega_bh^2=0.024$,
$\Omega_mh^2=0.14$, $h=0.72$, and $\tau=0.166$, where $\Omega_b$ and
$\Omega_m$ are density parameters of baryon and non-relativistic
matter, respectively, $h$ the Hubble constant in units of 100\ 
km/sec/Mpc, and $\tau$ the reionization optical depth, which are suggested 
from the WMAP experiment \cite{aph0302209}.  We also neglect the scale
dependence of $S_i^{(\delta\phi_A)}$, which is expected to be small in
a large class of inflationary models.) As one can see, for
non-vanishing value of $\kappa_m$, the shape of the CMB angular power
spectrum may significantly deviate from the result with the adiabatic
fluctuation.  Since the observed CMB angular power spectrum by the
WMAP experiment is highly consistent with the prediction from the
adiabatic density fluctuation, a stringent constraint on $\kappa_{\rm
iso}$ and $\kappa_{\rm corr}$ can be obtained.

\begin{figure}[t]
    \centering \includegraphics[width=10cm]{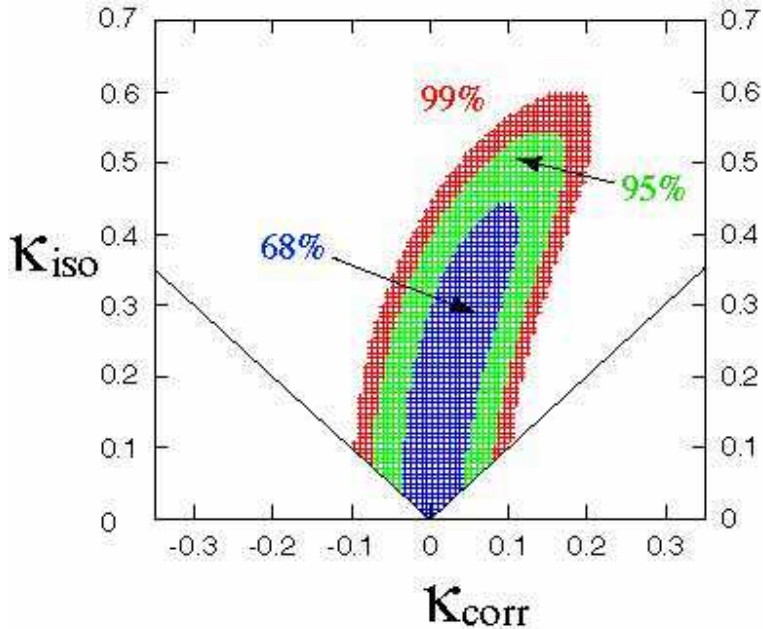}
    \caption{Probability contours for mixture of adiabatic and
    correlated isocurvature fluctuations. Note that we consider only the region
    of $\kappa_{\rm iso} \geq  \left|\kappa_{\rm corr}\right|$.}
    \label{fig:kappa-const}
\end{figure}

As one can expect from Fig.\ \ref{fig:kappa-dep}, if $\kappa_{\rm
iso}$ or $\kappa_{\rm corr}$ becomes too large, the CMB angular power
spectrum deviates from the adiabatic result.  In order to derive
constraints on these parameters, we perform likelihood analysis using
the WMAP data.  In our analysis, we first calculate the CMB angular
power spectrum for given values of $\kappa_{\rm iso}$ and $\kappa_{\rm
corr}$.  Then we calculate the likelihood variable using the numerical
program provided by the WMAP collaboration with the WMAP data
\cite{hinshaw}.  Assuming that the probability distribution is
proportional to the likelihood variable, we obtain the probability
distribution on the $\kappa_{\rm corr}$ vs. $\kappa_{\rm iso}$ plane,
where the total probability integrated over the plane with the uniform
measure is normalized to be unity (Bayesian analysis~\cite{PDG}).

The probability contours are shown in Fig.~\ref{fig:kappa-const}. 
From the figure, it is seen that $\kappa_{\rm iso}=0.4$ is allowed at 95\
\% C.L. for the pure (uncorrelated) isocurvature case.  We can also
see that a larger value of $\kappa_{\rm iso}$ is allowed if positive
correlation ($\kappa_{\rm corr}>0$) exists.  In the following
analysis, we derive constraints on various curvaton scenarios using
the result presented in Fig.\ \ref{fig:kappa-const}.

\section{Cosmological Moduli Fields as Curvaton}
\label{sec:moduli}
\setcounter{equation}{0}

We first consider the case where the moduli fields, which are flat
directions in the superstring theory, play the role of the curvaton.
Interactions of the moduli fields are expected to be proportional to
inverse power(s) of the gravitational scale, $M_*$, and their
potential is lifted only by the effect of the supersymmetry breaking.
If the moduli fields have non-vanishing amplitude in the early
universe, they might dominate the universe at a later stage and reheat
the universe when they decay.  Originally, such moduli fields are
considered to be cosmologically dangerous since their lifetimes may be
so long that they might decay after the big-bang nucleosynthesis
(BBN).  In this case, the decay products of the moduli fields would
spoil the success of the BBN \cite{ModuliProb}.  However, it was
pointed out that such a problem may be avoided if the lifetimes of the
moduli fields are shorter than $\sim 1\ {\rm sec}$ \cite{HeavyModuli}.
This is the case if the masses of the moduli fields are heavier than
about $10\ {\rm TeV}$ or so.  If such cosmological moduli fields
exist, they play the role of the curvaton if the primordial
fluctuations are generated in the amplitudes of the moduli fields
during inflation.

Let us first consider the isocurvature fluctuation in the baryonic
component.  In discussing the effects of the primordial fluctuation in
the curvaton, we here neglect the primordial fluctuation in the
baryonic sector, which may become independent and uncorrelated source
of the cosmic density fluctuation.  In order to parameterize the
isocurvature fluctuation in the baryonic sector, we define
\begin{eqnarray}
    \kappa_b^{(\delta\phi_A)} \equiv 
    \left[ \frac{S_{b\gamma}^{(\delta\phi_A)}}{\Psi^{(\delta\phi_A)}} 
    \right]_{\rm RD2}.
\end{eqnarray}

Fluctuation in the baryonic sector strongly depends on when the baryon
asymmetry of the universe is generated.  If the baryon asymmetry is
generated after the $\phi$D epoch is realized, 
primordial fluctuation in the curvaton is imprinted into the baryonic sector and the
isocurvature fluctuation in the baryonic component does not arise.
This happens when the baryon asymmetry is somehow generated from the
thermally produced particles after the decay of the curvaton fields.
In addition, if we consider the Affleck-Dine baryogenesis, $\kappa_b$
vanishes when the Affleck-Dine field starts to move after the beginning of
the $\phi$D epoch.  If the baryon asymmetry is generated 
before the $\phi$D epoch, on the contrary, 
fluctuation of the baryon density becomes negligibly small since
baryogenesis occurs when the metric perturbations
$\Psi^{(\delta\phi_A)}$ has not grown enough.  
%In this case,
%fluctuation of the baryon density becomes negligibly small.  
Even if there is no primordial fluctuation in the baryonic sector, however,
isocurvature fluctuation may arise in the baryonic component since
density fluctuation of $\hat{\phi}$ is inherited to the density
fluctuation of the radiation.  As we mentioned, $S_{i}$ becomes the
entropy between components generated from $\phi_i$ and those not from
$\phi_i$ and hence $S_{b\gamma}^{(\delta\phi_A)} = - \sum_i
f_{\gamma_i}
S_{i}^{(\delta\phi_A)}=\frac{9}{2}[\Psi^{(\delta\phi_A)}]_{\rm RD2}$.
In summary, the value of $\kappa_b$ depends on when the baryon
asymmetry of the universe is generated:
\begin{eqnarray}
    \kappa_b =
    \left\{ \begin{array}{ll}
            0 & ~~~:~~~ {\rm baryogenesis\ during/after\ {\phi}D} \\
            9/2 & ~~~:~~~ {\rm baryogenesis\ before\ {\phi}D}
        \end{array} \right. .
    \label{eq:kapbaryon}
\end{eqnarray}

Next we consider the isocurvature fluctuation in the CDM sector.  One
important point is that, since the interactions of the moduli fields
are very weak, it is natural to expect that the reheating temperature
due to the decay of the moduli fields is very low.  This fact has an
important implication to the scenario where the lightest
supersymmetric particle (LSP) becomes the CDM since the reheating
temperature may be lower than the freeze-out temperature of the LSP.
If so, they should be non-thermally produced to account for the CDM.
In this case fluctuations in the curvaton fields may generate extra
isocurvature fluctuations in the CDM sector as will be discussed
below.  Of course, there are other well-motivated candidates of the
CDM, like the axion, with which the isocurvature fluctuation in the
CDM sector vanishes.\footnote{The axion starts its oscillation
during/after $\phi$D.}

Let us consider the most non-trivial case where the decay products of
the moduli fields directly produce the LSP~\cite{NPB570-455}.  In this
case, the CDM density as well as the densities of the curvatons and
radiation are determined by the following Boltzmann equations:
\begin{eqnarray}
   \frac{dn_{\tilde{\chi}_i}}{dt} + 3 H n_{\tilde{\chi}_i}
   &=& \bar{N}_{\tilde{\chi},i} \Gamma_{\phi_i} n_{\phi_i}
   - \langle v_{\rm rel}\sigma\rangle 
   n_{\tilde{\chi}} n_{\tilde{\chi}_i},
   \label{dnchi/dt} \\
   \frac{dn_{\phi_i}}{dt} + 3Hn_{\phi_i} &=& 
   -\Gamma_{\phi_i} n_{\phi_i},
   \\
   \frac{d\rho_{\gamma_i}}{dt} + 4H\rho_{\gamma_i} &=&
   (m_{\phi_i} - \bar{N}_{\tilde{\chi},i} m_{\tilde{\chi}}) 
   \Gamma_{\phi_i} n_{\phi_i}
   + m_{\tilde{\chi}} \langle v_{\rm rel}\sigma\rangle 
   n_{\tilde{\chi}} n_{\tilde{\chi}_i},
   \label{drho/dt}
\end{eqnarray}
where $H\equiv \dot{a}/a$, $n_{\tilde{\chi}_i}$ is the number density
of the LSP produced from the decay of $\phi_i$ while
$n_{\tilde{\chi}}\equiv\sum_i n_{\tilde{\chi}_i}$ is the total number
density of the LSP, $m_{\tilde{\chi}}$ is the mass of the LSP,
$\bar{N}_{\tilde{\chi},i}$ is the averaged number of the LSP produced
in the decay of one modulus particle $\phi_i$, and $\langle v_{\rm
rel}\sigma\rangle$ is the thermally averaged annihilation cross
section of the LSP.  (In our numerical calculation, we assume that the
LSP is the neutral wino, as suggested by the simple anomaly mediated model.
We expect
that $\tilde{\chi}\tilde{\chi}\rightarrow W^+W^-$ is the dominant
annihilation mode and we use
\begin{eqnarray}
    \langle v_{\rm rel}\sigma\rangle
    = \frac{g_2^4}{2\pi} \frac{1}{m_\chi^2}
    \frac{(1-x_W)^{3/2}}{(2-x_W)^{2}},
\end{eqnarray}
where $x_W\equiv m_W^2/m_{\tilde{\chi}}^2$ with $m_W$ being the
$W$-boson mass, and $g_2$ is the gauge coupling constant of
SU(2)$_{\rm L}$.  Our conclusions are qualitatively unchanged even if
the LSP is not wino.)

First, we consider the case with only one (real) modulus field.  (For
the moment, we drop the index $i$.)  Number density of the relic LSP
is sensitive to the parameters $\bar{N}_{\chi}$ and $\langle v_{\rm
rel}\sigma\rangle$.  If $\bar{N}_{\chi}\ll 1$, the pair annihilation
of the LSP becomes ineffective and hence almost all the produced LSPs
survive.  Then,
\begin{eqnarray}
    [ n_{\tilde{\chi}} ]_{H\sim\Gamma_\phi}
    \sim \bar{N}_{\tilde{\chi}} 
    [ n_\phi ]_{H\sim\Gamma_\phi}
    \sim \frac{\bar{N}_{\tilde{\chi}}\Gamma_\phi^2M_*^2}{m_\phi},
\end{eqnarray}
where $n_\phi\equiv\rho_\phi/m_\phi$ being the number density of
$\phi$, leading to
\begin{eqnarray}
        \label{eq:rhocs}
    \frac{\rho_c}{s} \sim
    \left[ \frac{m_{\tilde{\chi}}n_{\tilde{\chi}}}{T^3} 
    \right]_{H\sim\Gamma_\phi}
    \sim \bar{N}_{\tilde{\chi}} \Gamma_\phi^{1/2} M_*^{1/2}    
     \frac{m_{\tilde{\chi}}}{m_\phi}.
  \end{eqnarray}
Parameterizing the decay rate of the modulus field as
\begin{eqnarray}
  \Gamma_\phi = \frac{\lambda_\phi^2}{2\pi} \frac{m_\phi^3}{M_*^2},
\end{eqnarray}
we obtain
\begin{eqnarray}
    \frac{\rho_c}{s} \simeq
    5.8\times 10^{-6}\ {\rm GeV}
    \times
    \bar{N}_{\tilde{\chi}} \lambda_\phi
    \left(\frac{m_{\tilde{\chi}}}{100\ {\rm GeV}}\right)
    \left(\frac{m_\phi}{100\ {\rm TeV}}\right)^{1/2}
    \left(\frac{g_*}{10.75}\right)^{-1/4}.
    \label{rho/s_0}
\end{eqnarray}
Importantly, in this case, the resultant LSP density is insensitive to
$\langle v_{\rm rel}\sigma\rangle$.  On the contrary, if
$\bar{N}_{\tilde{\chi}}$ becomes large enough, the pair annihilation of
the LSPs cannot be neglected and the number density of the LSP is 
given by
\begin{eqnarray}
    [ n_{\tilde{\chi}} ]_{H\sim\Gamma_\phi}
    \sim
    \left[ 
        \frac{H}{\langle v_{\rm rel}\sigma\rangle}
    \right]_{H\sim\Gamma_\phi}
    \sim
    \frac{\Gamma_\phi}{\langle v_{\rm rel}\sigma\rangle}.
\end{eqnarray}
Thus, the resultant LSP density is inversely proportional to
$\langle v_{\rm rel}\sigma\rangle$.

Now we discuss the density fluctuations in the multi-field case.  As
we discussed in the previous section, the metric perturbation is
generated associated with the primordial fluctuations of the moduli
fields.  In discussing the cosmic density fluctuations, it is also
important to understand the properties of the isocurvature
fluctuations.  In the RD2 epoch, behaviors of the density and metric
perturbations can be understood by using Eqs.\ (\ref{delta_T(RD2)})
and (\ref{Psi(RD2)}).  Isocurvature fluctuation in the CDM sector is,
on the other hand, obtained by solving Eqs.\ (\ref{dnchi/dt}) $-$
(\ref{drho/dt}).

In estimating the isocurvature fluctuation in the CDM sector, it is
important to note that $S_{1}$ and $S_{2}$ should be different to
realize a non-vanishing value of $S_{c\gamma}$.  If the ratio
$n_{\phi_1}/n_{\phi_2}$ does not fluctuate, the system consisting of
$\phi_1$ and $\phi_2$ can be regarded as a single fluid and hence the
isocurvature fluctuation should vanish.  This fact implies that, if
$S_{1}-S_{2}=0$, no isocurvature fluctuations can be induced as far as
the components generated from the decay products of $\phi_1$ and
$\phi_2$ are concerned.  Thus, the isocurvature fluctuation should be
proportional to $S_{12}=S_{1}-S_{2}$, and
\begin{eqnarray}
    S_{c\gamma}^{(\delta\phi_A)} = 
    \frac{\partial \ln [n_c/n_\gamma]_{\rm RD2}}
    {\partial \ln [n_{\phi_1}/n_{\phi_2}]_{\phi{\rm D}}}
    S_{12}^{(\delta\phi_A)}.
    \label{S_cg}
\end{eqnarray}
Defining 
\begin{eqnarray}
    \kappa_c^{(\delta\phi_A)} \equiv 
    \left[ \frac{S_{c\gamma}^{(\delta\phi_A)}}{\Psi^{(\delta\phi_A)}}
    \right]_{\rm RD2},
\end{eqnarray}
we obtain
\begin{eqnarray}
    \kappa_c^{(\delta\phi_A)} = -\frac{9}{2} 
    \left[ 
        \frac{\bar{S}^{(\delta\phi_A)}}{S_{12}^{(\delta\phi_A)}} 
      + \frac{1}{2} (f_{\gamma_1} - f_{\gamma_2}) 
    \right]^{-1}
    \frac{\partial \ln [n_c/n_\gamma]_{\rm RD2}}
    {\partial \ln [n_{\phi_1}/n_{\phi_2}]_{\phi{\rm D}}}.
    \label{kappa_c}
\end{eqnarray}
Notice that
\begin{eqnarray}
    \kappa_m^{(\delta\phi_A)}  = (\Omega_b\kappa_b^{(\delta\phi_A)}  +   
    \Omega_c\kappa_c^{(\delta\phi_A)} )/\Omega_m.
\end{eqnarray}

As one can see, $\kappa_c$, which parameterizes the size of the
correlated isocurvature fluctuation in the CDM sector, depends on
various parameters.  First, it depends on the parameters related to
the properties of the moduli fields, in particular $\Gamma_{\phi_i}$
and $\bar{N}_{\tilde{\chi},i}$.  In addition, $\kappa_c$ is sensitive
to the initial amplitudes of the moduli fields on which $f_{\gamma_i}$
depends.  Furthermore, $\kappa_c$ depends on the transfer matrix
${\cal T}$ defined in Eq.\ (\ref{DefofT}) which parameterizes how the
primordial fluctuations in the moduli fields $\delta\phi_A$ propagate
to the fluctuations of the mass eigenstates $\delta\phi_i$.  In
particular, ${\cal T}$ determines the ratio
$\bar{S}^{(\delta\phi_A)}/S_{12}^{(\delta\phi_A)}$.  Since the
interactions of the moduli fields are suppressed by inverse powers of
$M_*$, the potential of $\hat{\phi}$ is expected to be almost
parabolic if the amplitude of $\hat{\phi}$ is smaller than $M_*$. 
Then, ${\cal T}$ becomes diagonal (in some basis) with a
good approximation, hence $\delta\phi_1$ and $\delta\phi_2$ become
uncorrelated. In this case, $\bar{S}/S_{12}=\pm\frac{1}{2}$.

In our study, we numerically solve Eqs.\ (\ref{dnchi/dt}) $-$
(\ref{drho/dt}) to obtain $\kappa_c$ given in Eq.\ 
(\ref{kappa_c}). Although $\kappa_c^{(\delta \phi_A)}$ has complicated 
dependence on various parameters, we can understand its behavior for
several cases
if ${\cal T}$ is given by Eq.\ (\ref{T-diag}).  Let us consider three
extreme cases:

\begin{itemize}

\item[1.]  $\Gamma_{\phi_1}\ll\Gamma_{\phi_2}$,
    $\bar{N}_{\tilde{\chi},i}\ll 1$
    and $[\phi_1/\phi_2]_{\rm inf}{}^2\gg
    {\rm Max}\left[m_{\phi_2}^2/m_{\phi_1}^2, ~
    \bar{N}_{\tilde{\chi},2} m_{\phi_2}/
    \bar{N}_{\tilde{\chi},1} m_{\phi_1}\right]$:
    
    The decay products of $\phi_2$ is cosmologically negligible in
    this case, {\it i.e.}, $f_{\gamma_1} \simeq 1$ and $f_{c_1} \simeq
    1$. Therefore, no isocurvature fluctuations are generated from the
    fluctuation of $\phi_1$.  On the contrary, $\delta\phi_2$ results
    in $\delta n_{\tilde{\chi}_2}/n_{\tilde{\chi}_2}\simeq
    \delta\rho_{\gamma_2}/\rho_{\gamma_2}\simeq S_{2}$.  As a result,
    \begin{equation}
        \kappa_c^{(\delta\phi_1)}\simeq 0,~~~
        \kappa_c^{(\delta\phi_2)}\simeq  -\frac{9}{2} 
        \left( \frac{f_{c_2}}{f_{\gamma_2}} - \frac{3}{4} \right).
    \end{equation} 
    The ratio of the metric perturbations is given as
    \beq
    \left|\frac{\Psi^{(\delta\phi_2)}}{\Psi^{(\delta\phi_1)}}\right|_{\rm RD2} 
    = \frac{f_{\gamma_2}}{f_{\gamma_1}}   
    \left[\frac{\phi_1}{\phi_2}\right]_{\rm inf}
    \simeq 
    \left[\frac{\phi_2}{\phi_1}\right]_{\rm inf}
     \frac{m_{\phi_2}^2 \Gamma_{\phi_1}^{2/3}}{
    m_{\phi_1}^2 \Gamma_{\phi_2}^{2/3}}\ll1,
    \eeq
    where it should be noted that the masses of the moduli fields are
    of the same order of magnitude as mentioned in Sec.~\ref{sec:framework}.
    Thus $\delta \phi_2$ does not contribute to $\kappa_{\rm corr}$ nor
    $\kappa_{\rm iso}$ as long as $\bar{N}_{\tilde{\chi},1} \gtrsim
     \bar{N}_{\tilde{\chi},2}$, even though $\kappa_c^{(\delta\phi_2)}$ is finite
    (see Eq.~(\ref{eq:relkappa})). Thus we expect that the density
    fluctuation is purely adiabatic in this case. On the other hand, if
    $\left[\phi_1/\phi_2\right]_{\rm inf} \ll  \bar{N}_{\tilde{\chi},2} m_{\phi_2}/
     \bar{N}_{\tilde{\chi},1}  m_{\phi_1}$ is satisfied, 
     $\kappa_{\rm iso}$ can be large due to $\delta \phi_2$, so the isocurvature
    fluctuation becomes substantial.   
    
\item[2.]  $\Gamma_{\phi_1}\gg\Gamma_{\phi_2}$,
    $\bar{N}_{\tilde{\chi},i}\ll 1$ and $[\phi_1/\phi_2]_{\rm
    inf}{}^2 \gtrsim  \bar{N}_{\tilde{\chi},2} m_{\phi_2}/
    \bar{N}_{\tilde{\chi},1} m_{\phi_1}$:
    
    First we focus on the case of $[\phi_1/\phi_2]_{\rm inf}{}^2\gg
    %{\rm Max}\left[
    \sqrt{m_{\phi_2}^4 \Gamma_{\phi_1}/
    m_{\phi_1}^4 \Gamma_{\phi_2}}%, ~
    %\bar{N}_{\tilde{\chi},2}/\bar{N}_{\tilde{\chi},1}\right]
    $.  In
    this case, most of the radiation and the CDM are generated from
    the decay products of $\phi_1$, and hence $f_{\gamma_1}\simeq 1$
    and $f_{c_1}\simeq 1$.  Then we obtain the same result as in the
    previous case;
    \begin{eqnarray}
        \kappa_c^{(\delta\phi_1)}\simeq 0,~~~
        \kappa_c^{(\delta\phi_2)}\simeq 
        -\frac{9}{2} 
        \left( \frac{f_{c_2}}{f_{\gamma_2}} - \frac{3}{4} \right).
    \end{eqnarray}
    It is more interesting to consider the case of
    $ [\phi_1/\phi_2]_{\rm inf} {}^2 \lesssim \sqrt{m_{\phi_2}^4\Gamma_{\phi_1}/
    m_{\phi_1}^4 \Gamma_{\phi_2}}$. Then, most of
    the radiation comes from the decay of $\phi_2$, while most of the
    CDM are generated from $\phi_1$, namely, $f_{\gamma_2}\simeq 1$
    and $f_{c_1}\simeq 1$.  Thus we expect large isocurvature
    fluctuation.
        
\item[3.] $\Gamma_{\phi_1}=\Gamma_{\phi_2}$ and
    $\bar{N}_{\tilde{\chi},i}\ll 1$:
    
    In this case, $\phi_1$ and $\phi_2$ simultaneously decay.
    Assuming $\bar{N}_{\tilde{\chi},i}\ll 1$, almost all of the CDMs
    produced by the decay do not experience the pair annihilation and
    hence the resultant number of the LSP is approximately
    proportional to $[\bar{N}_{\tilde{\chi},1}n_{\phi_1}+
    \bar{N}_{\tilde{\chi},2}n_{\phi_2}]_{H\sim\Gamma_\phi}$.  Using
    the fact that $[m_{\phi_1} n_{\phi_1}+
    m_{\phi_2}n_{\phi_2}]_{H\sim\Gamma_\phi}$ is
    fixed, we obtain
    \begin{eqnarray}
        \kappa_c^{(\delta\phi_1)} \simeq
        -\frac{9}{2f_{\gamma_1}} 
     %   \frac{2(\bar{N}_{\tilde{\chi},1}-\bar{N}_{\tilde{\chi},2})x}
     %   {(\bar{N}_{\tilde{\chi},1}+\bar{N}_{\tilde{\chi},2})(1+x)^2
     %   +(\bar{N}_{\tilde{\chi},1}-\bar{N}_{\tilde{\chi},2})(1-x^2)},
     	\frac{(\bar{N}_{\tilde{\chi},1} \,y-
	\bar{N}_{\tilde{\chi},2} x)}{(\bar{N}_{\tilde{\chi},1}+\bar{N}_{\tilde{\chi},2} \,x)
	(1+y)}
    \end{eqnarray}
    where $x \equiv [n_{\phi_2}/n_{\phi_1}]_{\rm inf}$ and 
    $y \equiv [m_{\phi_2} n_{\phi_2}/m_{\phi_1} n_{\phi_1}]_{\rm inf}$.
    $\kappa_c^{(\delta\phi_2)}$ is obtained by replacing
    $1\leftrightarrow 2$ and $x,\,y\rightarrow 1/x,\,1/y$.  Notice that the
    isocurvature fluctuation vanishes if
    $\bar{N}_{\tilde{\chi},1}/m_{\phi_1}=\bar{N}_{\tilde{\chi},2}/m_{\phi_2}$. 
    This is from the fact that $\phi_1$ and
    $\phi_2$ can be regarded as a single fluid, since 
   $n_{\tilde{\chi}_1}/ n_{\gamma_1} =  n_{\tilde{\chi}_2}/ n_{\gamma_2}$
   when  $\bar{N}_{\tilde{\chi},1}/m_{\phi_1}=\bar{N}_{\tilde{\chi},2}/m_{\phi_2}$.
   
\end{itemize}

\begin{figure}[t]
    \centering
    \includegraphics[width=7cm]{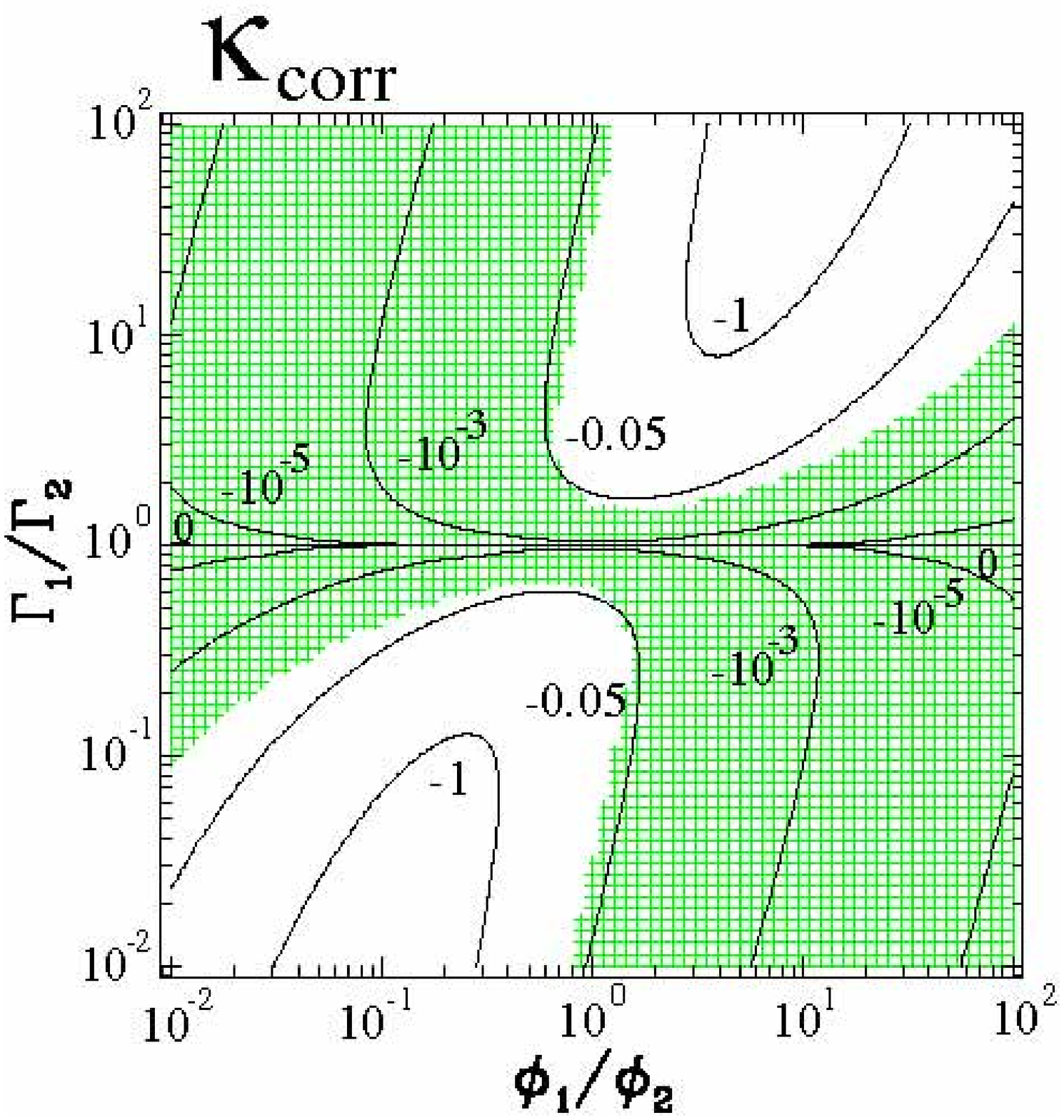}  
    \includegraphics[width=7cm]{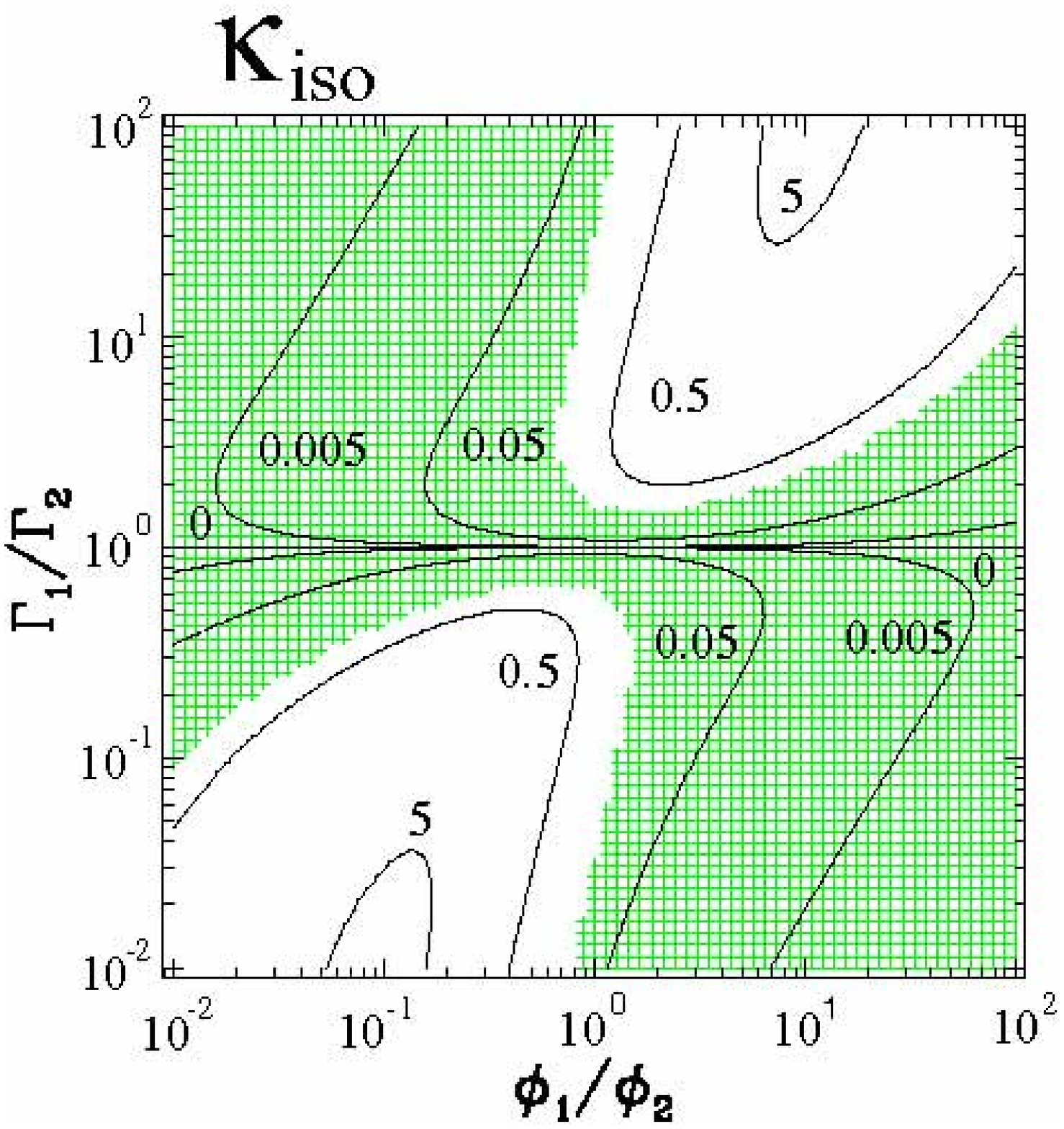}
    \caption{%Probability contours 
    Contours for  the values of $\kappa_{\rm iso}$ and 
    $\kappa_{\rm corr}$ in $\Gamma_1/\Gamma_2$ vs.\
    $[\phi_1/\phi_2]_{\rm inf}$ plane for $N_{\tilde{\chi},1}=10^{-3}$,
    $N_{\tilde{\chi},2}= 10^{-3}$ and $\kappa_b=0$. We also show the 
    allowed area within 95 \% C.L. as the shaded (green) region.
    Here $m_{\phi_1}= m_{\phi_2}$ is assumed.
     }
    \label{fig:moduli1}
\end{figure}

\begin{figure}[t]
    \centering
    \includegraphics[width=7cm]{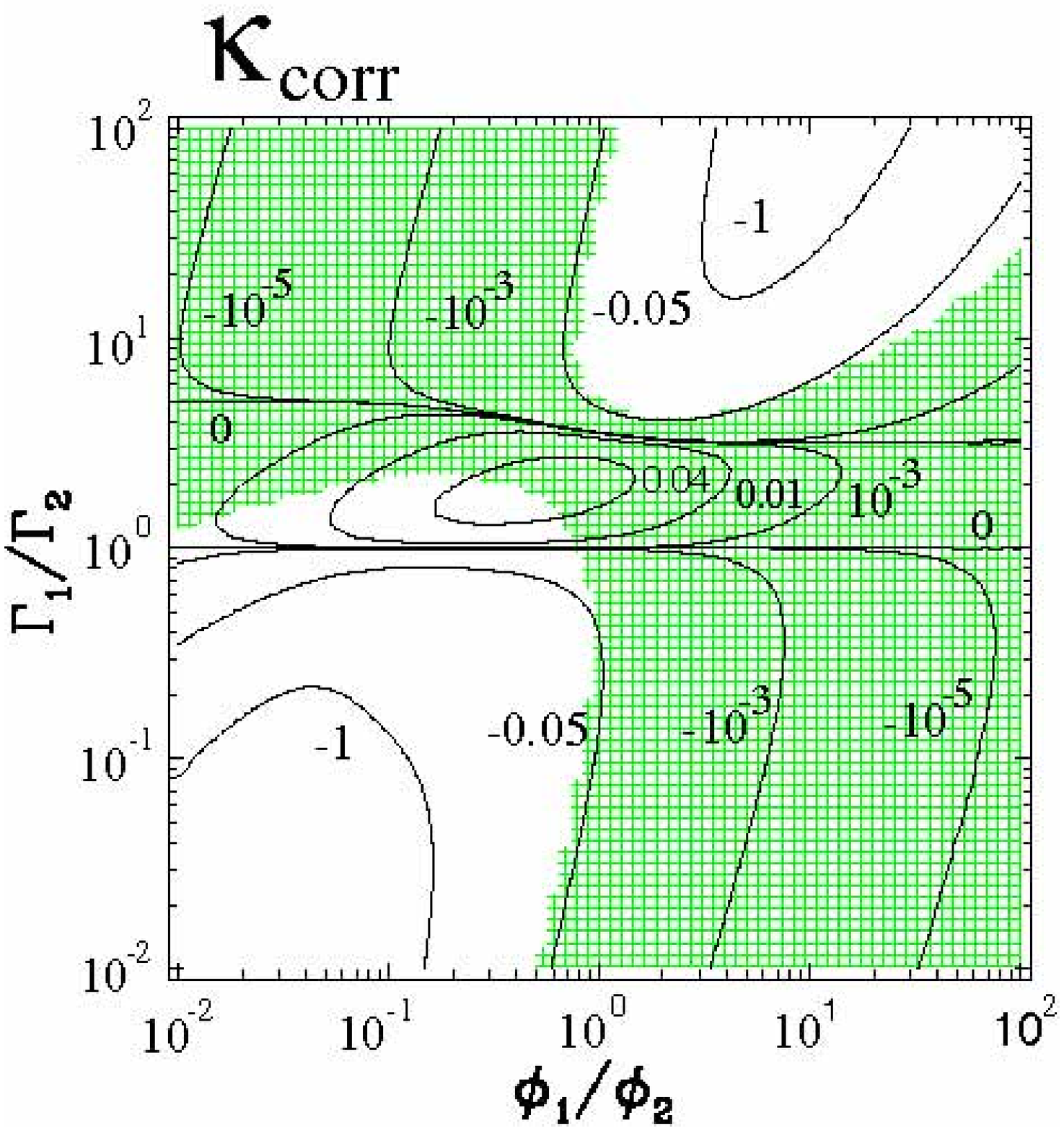} 
    \includegraphics[width=7cm]{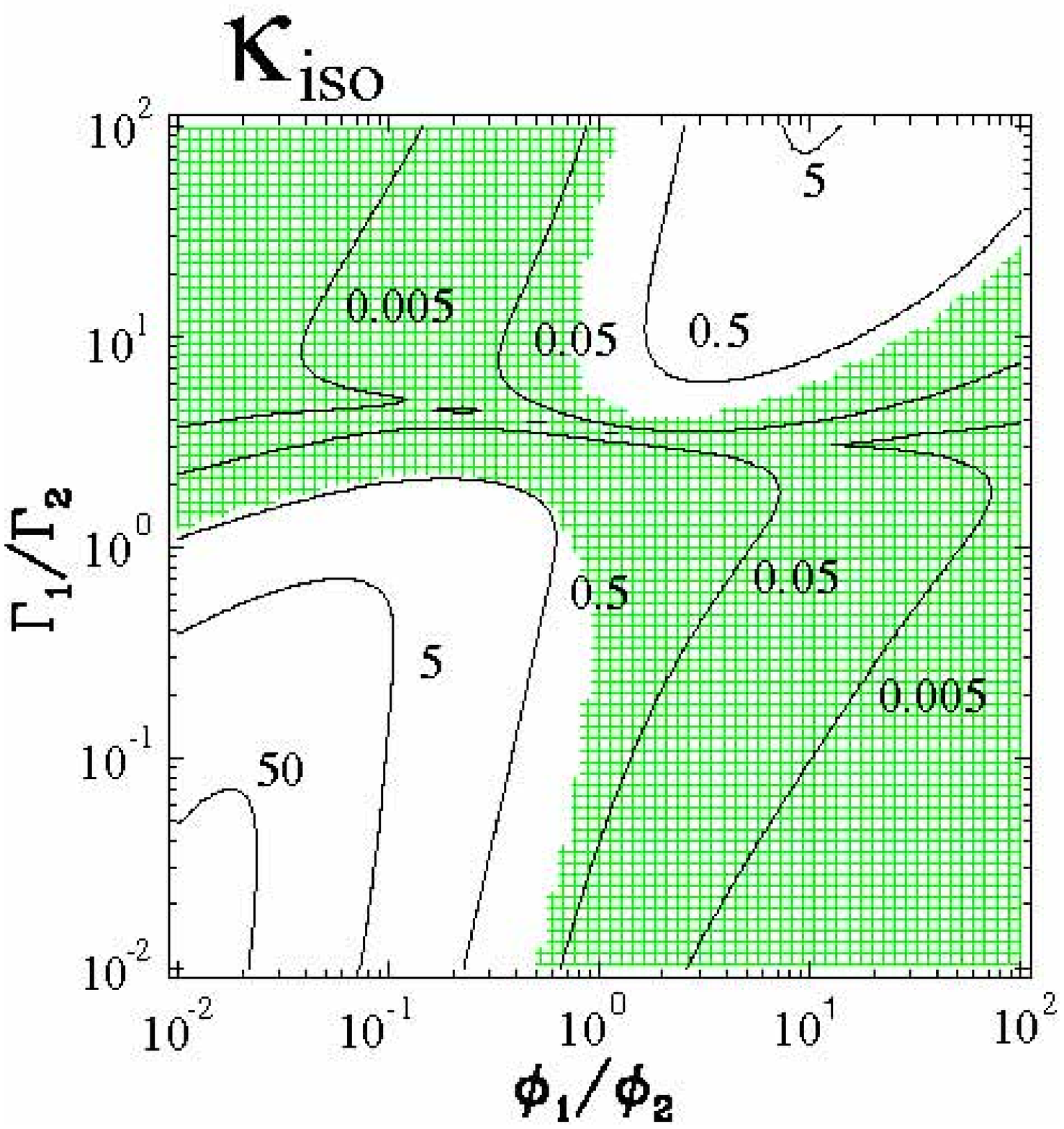}
    \caption{Same as Fig.~\ref{fig:moduli1} except 
    $N_{\tilde{\chi},1}= 0.1$ and $N_{\tilde{\chi},2}=10^{-3} $. }
    \label{fig:moduli2}
\end{figure}

We numerically calculate $\kappa_{\rm corr}$ and $\kappa_{\rm iso}$
with the use of Eq.\ (\ref{eq:relkappa}). Contours of constant
$\kappa_{\rm corr}$ and $\kappa_{\rm iso}$ are shown in Figs.\
\ref{fig:moduli1} and \ref{fig:moduli2}.  Then, with $\kappa_{\rm
corr}$ and $\kappa_{\rm iso}$, the CMB angular power spectrum is
calculated and we derive constraints on the curvaton model under
consideration making use of the likelihood analysis as Fig.\
\ref{fig:kappa-const}. In the numerical calculation and  the
discussion hereafter, we assume $m_{\phi_1}= m_{\phi_2}$ for simplicity.
Also we take $m_{\chi}=100$ GeV, and $\lambda_{\phi_i}^2$ runs from $0.1$
to $10$. 
\footnote
{In our numerical calculation, we have tuned the mass of the moduli
fields so that the pair annihilation cross section of the lightest
neutralino realizes our canonical value of the relic density of the
CDM. 
}

We first consider the case with $\kappa_b=0$.  Even in this case,
there are two scalar fields $\phi_1$ and $\phi_2$ and hence, in
general, correlated mixture of the adiabatic and isocurvature
fluctuations may arise.  In order to discuss the effects of the
correlated isocurvature fluctuation, we calculate the likelihood
variable and derive the constraint.  In Figs.\ \ref{fig:moduli1} and
\ref{fig:moduli2}, we plot the contours for the probabilities obtained
from the likelihood analysis.  The results with $N_{\tilde{\chi},1} =
N_{\tilde{\chi},2} = 10^{-3}$ (see Fig.~\ref{fig:moduli1}) can be
relatively easily understood.  In this case, almost all the components
are generated from $\phi_1$ when $\Gamma_1/\Gamma_2\ll 1$ and
$[\phi_1/\phi_2]_{\rm inf}\gg 1$ ({\it i.e.}, the lower-right corner
of the figures), while the decay products of $\phi_2$ generates almost
all the components when $\Gamma_1/\Gamma_2\gg 1$ and
$[\phi_1/\phi_2]_{\rm inf}\ll 1$ ({\it i.e.}, the upper-left corner of
the figures).  In these cases, the situation is like the case with
single scalar field and hence the CMB angular power spectrum almost
agrees with the adiabatic one.  Remarkably, in order to realize the
``adiabatic-like'' result, the ratios $\Gamma_1/\Gamma_2$ and
$[\phi_1/\phi_2]_{\rm inf}$ do not have to be extremely large as can
be seen in Fig.~\ref{fig:moduli1}. In the upper-right and lower-left
corners of the figures, decay products of both $\phi_1$ and $\phi_2$
produce significant amount of the radiation and/or the CDM. Note that
the second extreme case mentioned before cannot be seen in this
figure, because $N_{\tilde{\chi},1} = N_{\tilde{\chi},2} = 10^{-3}$
are not small enough and hence the pair annihilation of the LSP is not
negligible.  
%In fact, the peaks of $|\kappa_{\rm corr}|$ and
%$\kappa_{\rm iso}$ correspond to the case that the amount of the CDM
%from $\phi_1$ is comparable to that from $\phi_2$. 
We have found that $|\kappa_{\rm corr}|$ and $\kappa_{\rm iso}$ are
enhanced when the amount of the CDM from $\phi_1$ is comparable
to that  from $\phi_2$. In these cases,
$S_{c\gamma}$ becomes sizable and the shape of the CMB angular power
spectrum becomes different from that from the adiabatic fluctuations.
Indeed, the CMB angular power spectrum at lower multipoles is enhanced
relative to that at higher multipoles due to this effect. Furthermore,
as pointed out above, one can see that the isocurvature fluctuation
vanishes along the horizontal axis with $\Gamma_1 = \Gamma_2$.  Even
if there exists large hierarchy between $N_{\tilde{\chi},1}$ and
$N_{\tilde{\chi},2}$, the constraint does not change much.  In
Fig.~\ref{fig:moduli2}, we plot the result for the case
$N_{\tilde{\chi},1}\gg N_{\tilde{\chi},2}$.

When $\kappa_b\neq 0$, the constraint completely changes.  As we
mentioned, $\kappa_b=9/2$ if the baryon asymmetry is generated before
the $\phi$D epoch, assuming vanishing primordial fluctuation in the
baryonic component.  In this case,  the correlated isocurvature fluctuation
 almost always becomes large, leading to
unacceptably large $\kappa_{\rm corr}$ and $\kappa_{\rm iso}$
irrespective of the choice of $N_{\tilde{\chi},1}$ and
$N_{\tilde{\chi},2}$.  Indeed, in this case, we have found that
the predicted CMB angular power spectrum becomes consistent with the
WMAP result in a very tiny parameter region.

Thus we conclude that mild hierarchy in $\Gamma_1/\Gamma_2$ is
necessary to realize the CMB angular power spectrum consistent with
the WMAP results, in the case of vanishing baryonic isocurvature
fluctuation. On the other hand, if there is nonzero baryonic
isocurvature fluctuation, it is difficult to obtain the CMB angular
power spectrum consistent with the observations.  In order to have
successful curvaton scenario, we need to suppress $\kappa_b$ as
$\kappa_b \lesssim 1$, which would constrain the scenario of
baryogenesis.

%\rem{Condition to suppress $\kappa_c$: vanishing $S_{12}$ with
%non-vanishing $\bar{S}$.}

\section{Affleck-Dine Field as Curvaton}
\label{sec:ad}
\setcounter{equation}{0}

Another class of possible curvaton fields in supersymmetric models includes
$F$- and $D$-flat directions.  There are various flat directions in
the supersymmetric models, which are lifted only by the effect of the
supersymmetry breaking.  One important application of such a flat
direction to cosmology is the Affleck-Dine baryogenesis where baryon
asymmetry of the universe originates from the condensation of such a
flat direction \cite{NPB249-361}.  Thus, now we consider the case
where the curvaton field is one of such flat directions.  In
particular, in this section, we study the possibility that the
Affleck-Dine field plays the role of the curvaton.  In this case, the
curvaton field $\hat{\phi}$ is also responsible for the baryon
asymmetry of the universe.  Another case will be discussed in the next
section.

In the Affleck-Dine scenario, the baryon asymmetry of the universe is
generated in the form of the coherent oscillation of the Affleck-Dine
field.  To make our discussion clearer, we adopt the following form of
the potential
\begin{eqnarray}
    V = m_\phi^2 |\hat{\phi}|^2 +
    \frac{\lambda m_\phi^2}{M_*^{n-2}} (\hat{\phi}^n + {\rm h.c.}),
\label{V(AD),noRG}
\end{eqnarray}
where the scalar field $\hat{\phi}$ has non-vanishing baryon number
$B_\phi$.  Here, since $\hat{\phi}$ is an $F$- and $D$-flat direction,
$m_\phi$ is induced by the effect of the supersymmetry breaking and is
of order the gravitino mass.\footnote
{In this section we assume the gravity-mediation. If the gravitino
mass $m_{3/2}$ is much larger than the soft mass $m_\phi$, as in the
case of anomaly-mediation \cite{NPB557-79,JHEP9812-027,JHEP0004-009},
the right-hand sides of Eqs.\ (\ref{n_B(H=m)}) and
(\ref{epsilon-approx}) are multiplied by a factor of
$(m_{3/2}/m_\phi)^2$. 
(In this case $\phi_{\rm init} \lesssim
M_*(m_\phi/m_{3/2})^{2/(n-2)}$ is required to avoid a
color/electromagnetic breaking minimum \cite{Fujii-Hamaguchi}.)
However, the conclusion of this section does not change.  We do not
consider the case of gauge-mediation \cite{GMSB}, where in general a
stable $Q$-ball is generated \cite{GMSB-Q}.}
In addition, $\lambda$ is a dimensionless coupling constant (which is
assumed to be $O(1)$) and $n$ is an integer larger than $2$.  (We
choose the coefficient of the higher-dimensional operator being real
using the phase rotation of $\hat{\phi}$.)  We expect that the
baryon-number violating term is induced from a higher dimensional
K\"ahler interaction and hence is proportional to the parameter
$m_\phi^2$.\footnote
{This is not the case if there is a non-renormalizable operator in the
superpotential. In such a case, the higher order term in Eq.\ 
(\ref{V(AD),noRG}) is proportional to $m_\phi$, and Eqs.\ 
(\ref{n_B(H=m)}) and (\ref{epsilon-approx}) are multiplied by
$(M_*/m_\phi)$.  However, the conclusion of this section does not
change, as long as the Affleck-Dine field plays the role of the
curvaton.}
As discussed in Section \ref{sec:perturbations}, we assume that the
potential is not affected by the Hubble-induced interactions.

The equation of motion of the Affleck-Dine field is given by
\begin{eqnarray}
   \ddot{\hat{\phi}} + 3 H \dot{\hat{\phi}} 
   + \frac{\partial V}{\partial \hat{\phi}^*} = 0,
\end{eqnarray}
which leads to
\begin{eqnarray}
   \dot{n}_B + 3 H n_B = i B_\phi
   \left( \frac{\partial V}{\partial \hat{\phi}} \hat{\phi}
   - {\rm h.c.} \right),
\end{eqnarray}
where the baryon-number density in this case is given by $n_B\equiv
iB_\phi(\hat{\phi}^*\dot{\hat{\phi}}-\dot{\hat{\phi}}^*\hat{\phi})$.
Assuming non-vanishing initial amplitude, nonzero baryon number is
generated when $\hat{\phi}$ starts to oscillate. The baryon number
density at that moment is estimated as
\begin{eqnarray}
   \left[ n_B \right]_{H\sim m_\phi} \sim
   \left[ B_\phi {\rm Im} 
   \left( \frac{\partial V}{\partial \hat{\phi}} \hat{\phi} \right)
   H^{-1} \right]_{H\sim m_\phi} 
   \sim
   B_\phi \frac{\lambda m_\phi}{M_*^{n-2}} 
   \phi_{\rm init}^n \sin (n\theta_{\rm init}),
   \label{n_B(H=m)}
\end{eqnarray}
where we have neglected coefficients of $O(1)$. Here, we have used the
approximation that the effect of the baryon-number violating operator
is so small that it can be treated as a perturbation.  (Validity of
such an approximation will be shown later.)  As one can see, the
baryon-number density becomes smaller as the initial amplitude of the
Affleck-Dine field is suppressed.

The energy density of the Affleck-Dine field is given by
$\rho_\phi\simeq m_\phi^2|\hat{\phi}|^2$, where we neglected the
effect of the higher dimensional operator assuming $\phi_{\rm init}$
is small enough.  Thus, defining
\begin{eqnarray}
   \epsilon \equiv \left[ \frac{m_\phi n_B}{\rho_\phi} 
   \right]_{\phi{\rm D}},
   \label{epsilon}
\end{eqnarray}
the $\epsilon$ parameter is given by 
\begin{eqnarray}
   \epsilon \sim B_\phi \lambda
   \left( \frac{\phi_{\rm init}}{M_*} \right)^{n-2}
   \sin (n\theta_{\rm init}).
   \label{epsilon-approx}
\end{eqnarray}
Importantly, when $H\lesssim m_\phi$, $n_B$ and $\rho_\phi$ are both
proportional to $a^{-3}$ and hence the ratio $n_B/\rho_\phi$ 
is a constant of
time as far as $H\gtrsim\Gamma_\phi$.  (In the case of $Q$-ball
formation, $\rho_\phi$ and $\Gamma_\phi$ should be understood as the
energy density and the decay rate of the $Q$-ball, respectively.)

When the Affleck-Dine field oscillates fast enough with the parabolic
potential, the motion of the Affleck-Dine field is described as
\begin{eqnarray}
   \phi_1 = A_1 (t) \cos (m_\phi t),~~~
   \phi_2 = A_2 (t) \sin (m_\phi t),
   \label{eq:A1A2}
\end{eqnarray}
where we have used the phase rotation of the Affleck-Dine field to
obtain the above form with $A_1\geq A_2$.  The amplitudes $A_{1}(t)$
and $A_{2}(t)$ are proportional to $a^{-3/2}$ when $\Gamma_\phi\ll
H\ll m_\phi$.

Evaluating the ratio $n_B/\rho_\phi$ at the time of the decay of the
Affleck-Dine field (or the $Q$-ball), the resultant baryon-to-entropy
ratio is estimated as
\begin{eqnarray}
   \frac{n_B}{s} \sim \epsilon \frac{T_{\rm RD2}}{m_\phi},
   \label{n_B/s}
\end{eqnarray}
where $T_{\rm RD2}$ is the reheating temperature due to the curvaton
decay.  In order not to spoil the success of the big-bang
nucleosynthesis, $T_{\rm RD2}$ should be higher than about $1\ {\rm
MeV}$~\cite{PRL82-4168}.  In addition, $m_\phi$ is as large as the
gravitino mass and is expected to be smaller than $\sim 1\ {\rm TeV}$
assuming the supersymmetry as a solution to the naturalness problem in
the standard model.  Consequently, to explain the presently observed
value of the baryon-to-entropy ratio of $n_B/s\sim O(10^{-10})$, 
the $\epsilon$ parameter should be much
smaller than $1$, that is, $\epsilon \lesssim 10^{-4}$.  This fact
means that a large hierarchy between $A_1$ and $A_2$ is required:
$|A_2/A_1|\lesssim 10^{-4}$.  Such a hierarchy is realized when, for
example, the initial amplitude of the Affleck-Dine field is smaller
than the suppression scale of the baryon-number violating higher
dimensional operator ({\it i.e.}, in our example, $M_*$).  The above
fact tells us that the motion of the curvaton field is almost in the
radial direction.  In addition, the field $\phi_r$ and $\phi_\theta$
(almost) correspond to the mass eigenstates $\phi_1$ and $\phi_2$,
respectively, since the scalar potential is well approximated by the
parabolic one.  As we will discuss, primordial fluctuations in
$\phi_r$ and $\phi_\theta$ give rise to different type of density
fluctuations.

Importantly, behaviors of the Affleck-Dine field strongly depends on
the detailed shape of the potential.  Even when the amplitude of
$\hat{\phi}$ is small enough so that the higher dimensional
interaction in Eq.\ (\ref{V(AD),noRG}) can be disregarded, the scalar
potential may deviate from the simple parabolic form once the
renormalization group effects are taken into account:
\begin{eqnarray}
    V = m_\phi^2 (M_*) 
    \left[ 1 + K \log ( |\hat{\phi}|^2 / M_*^2 ) \right] 
    |\hat{\phi}|^2,
    \label{V(AD),withRG}
\end{eqnarray}
where we chose $M_*$ as a renormalization point.  The parameter $K$ is
from the renormalization group effect.  If the Yukawa interaction
associated with $\hat{\phi}$ is weak enough, the gauge interaction
determines the scale dependence of $m_\phi$ and the parameter $K$ then
becomes negative.  If the Yukawa interaction becomes strong, on the
contrary, $m_\phi^2$ is more enhanced at higher energy scale and the
parameter $K$ becomes positive.  Thus, if the Affleck-Dine mechanism
works with the flat direction consisting of first and/or second generation MSSM
particles, the parameter $K$ is likely to be negative.  If a third
generation squark (in particular, stop) or the up-type Higgs field is
associated with the flat direction, however, $K$ may become positive.

Evolution of the coherent oscillation depends on the sign of $K$.  If
the potential of $\hat{\phi}$ is flatter than the parabolic one,
coherent oscillation of the scalar field $\hat{\phi}$ may evolve into
the non-topological solitonic objects called ``$Q$-ball'' (or in our
case, ``$B$-ball'') \cite{Q-ball}.  Properties of the density
fluctuations differ for the cases with and without the $Q$-ball
formation.  Thus, we discuss these two cases separately.

\subsection{Case without $Q$-ball formation}
\label{sec:woQ}

First, we consider the case where the $Q$-balls are not formed.  This
is the case when the parameter $K$ is positive.  In this case, it is
expected that the reheating temperature after the decay of the
Affleck-Dine field is relatively high, {\it i.e.}, higher than the
freeze-out temperature of the lightest neutralino which is assumed to
be the CDM.  In this case, the relic LSPs are thermally produced and
relic density of the LSP (more precisely, $n_c/s$) becomes independent
of the reheating temperature due to the decay of the Affleck-Dine
field.  Thus, in this case, no entropy fluctuation between the CDM and
the radiation is generated.  

%On the other hand, if the CDM is not the
%LSP, the isocurvature fluctuation between the CDM and the radiation
%may arise. For example, consider the axion CDM. Then the fluctuation
%of the axion field results in the pure isocurvature fluctuation, while
%the correlated one is not generated. Hereafter, however, we just
%neglect the entropy fluctuation in the CDM sector, since the following
%discussion would suffice for turning down the present case.
%\rem{Stataement unclear????}

%\rem{Case where the CDM is not LSP?}

As we mentioned, when the amplitude of the Affleck-Dine field is much
smaller than the suppression scale of the baryon-number violating
higher dimensional interactions, the fields $\phi_r$ and $\phi_\theta$
(almost) correspond to the mass eigenstates $\phi_1$ and $\phi_2$,
respectively.  Primordial fluctuations in $\phi_r$ and $\phi_\theta$
result in different type of density fluctuations.  Because of the
large hierarchy between the amplitude of the fields
$\phi_1\simeq\phi_r$ and $\phi_2\simeq\phi_\theta$, almost all the
photons in the RD2 epoch are generated from the decay products of
$\phi_1$ and hence $f_{\gamma_1}\simeq 1$ while $f_{\gamma_2}\simeq
0$.  Furthermore, since the Affleck-Dine field mostly feels the
parabolic potential, the primordial entropy fluctuations are given by
\begin{eqnarray}
   S_{i} = \left[ \frac{2\delta\phi_i}{\phi_i} \right]_{\rm inf}.
\end{eqnarray}
As a result, using Eq.\ (\ref{Psi(RD2)}), we obtain\footnote
{In fact, associated with $\delta\phi_\theta$, the metric perturbation
$\left[ \Psi^{(\delta\phi_\theta)} \right]_{\rm RD2}$ may be
generated.  However, it is suppressed by a factor of
$f_{\gamma_2}S_2/f_{\gamma_1}S_1\sim
|A_2/A_1|(\delta\phi_\theta/\delta\phi_r)\lesssim 10^{-4}$, compared with
$\left[ \Psi^{(\delta\phi_r)} \right]_{\rm RD2}$.  Metric perturbation
of this order does not significantly change the shape of the CMB power
spectrum and we neglect such an effect.}
\begin{eqnarray}
    \left[ \Psi^{(\delta\phi_r)} \right]_{\rm RD2} \simeq 
    -\frac{4}{9} \frac{\delta\phi_r}{\phi_{\rm  init}},~~~
    \left[ \Psi^{(\delta\phi_\theta)} \right]_{\rm RD2} \simeq 0.
\end{eqnarray}
In addition, since $\delta\phi_r$ and $\delta\phi_\theta$ provide
fluctuations in $\phi_{\rm init}$ and $\theta$, respectively, the entropy
fluctuation between the baryon and radiation is also expected.
Indeed, using the fact that the $\epsilon$ parameter given in Eq.\
(\ref{epsilon}) is proportional to the resultant baryon-to-entropy
ratio, entropy fluctuation in the baryonic sector is estimated:
\begin{eqnarray}
   S_{b\gamma}^{(\delta\phi_r)} = 
   (n-2) \frac{\delta\phi_r}{\phi_{\rm  init}},~~~
   S_{b\gamma}^{(\delta\phi_\theta)} = 
   n \cot (n\theta_{\rm init}) 
   \frac{\delta\phi_\theta}{\phi_{\rm  init}}.
   \label{eq:Sbgamma}
\end{eqnarray}

\begin{figure}
    \centering
    \includegraphics[width=10cm]{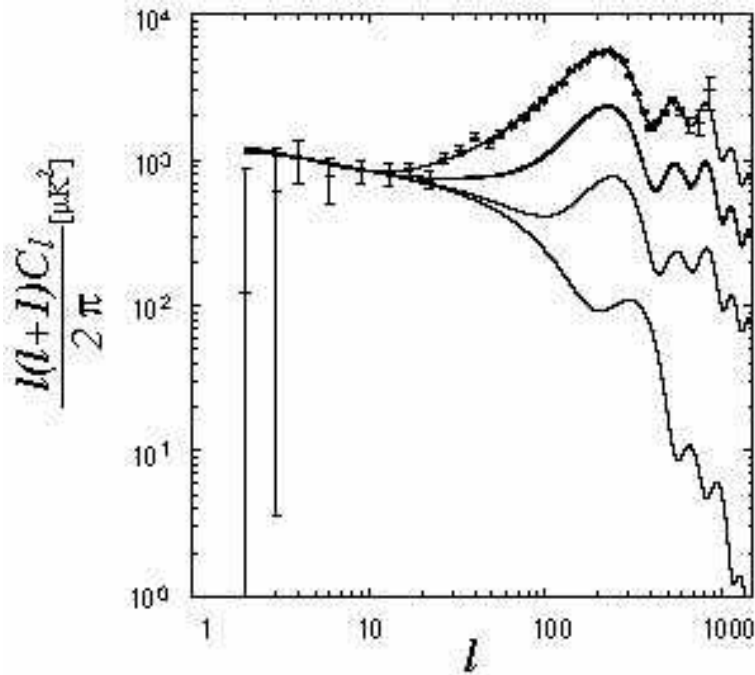}
    \caption{The predicted CMB angular power spectra for the case
    where the Affleck-Dine field plays the role of the curvaton. 
    From top to bottom, the WMAP data and the best-fit $\Lambda$CDM
    model, the case without  the $Q$-ball formation,
    the case with  the $Q$-ball formation ($C=1$) which  accidentally 
    coincides with the previous case, the case 
    with  the $Q$-ball formation ($C=0$), and  the purely isocurvature case.
%    The dotted (blue) line and dashed (pink) line show the cases
%    without and with the $Q$-ball formation, respectively.
    We take $n=4$, $\cot n\theta_{\rm init}=0$ and normalize the
    Sachs-Wolfe tail. 
    % The WMAP data, the best-fit $\Lambda$CDM
   % model (solid (red) line) and     
    %the result for purely isocurvature case (dashed and single-dotted (light blue) 
    %line) are also shown.
    %\rem{Takahashi-kun, please check and modify!!!}
     }
    \label{fig:ad_cmb}
\end{figure}

Since the fluctuations $\delta\phi_r$ and $\delta\phi_\theta$ are
uncorrelated, the total CMB angular power spectrum is given by a
linear combination of the two power spectra $C_l^{(\delta\phi_r)}$ and
$C_l^{(\delta\phi_\theta)}$, which are characterized by the following
$\kappa_m$ parameters:
\begin{eqnarray}
    \kappa_m^{(\delta\phi_r)} \simeq -\frac{9}{4} 
    \frac{(n-2)\Omega_b}{\Omega_m},
    ~~~ \kappa_m^{(\delta\phi_\theta)} \simeq
    \infty.
    \label{eq:kappam}
\end{eqnarray}
Thus, the CMB angular power spectrum associated with
$\delta\phi_\theta$ is (almost) the same as the purely isocurvature
result while $C_l^{(\delta\phi_r)}$ is from the correlated mixture of
the adiabatic and isocurvature fluctuations.  Relative size of
$C_l^{(\delta\phi_r)}$ and $C_l^{(\delta\phi_\theta)}$ depends on
$\theta_{\rm init}$.  Since $\kappa_m^{(\delta\phi_r)}<0$, however,
the acoustic peaks are always suppressed compared to the adiabatic
case once one normalizes the Sachs-Wolfe tail.  Since the angular
power spectrum measured by the WMAP is well consistent with the
adiabatic result, suppression of the acoustic peaks causes discrepancy
between the theoretical prediction and the observations.  To study
this issue, in Fig.\ \ref{fig:ad_cmb}, we plot the predicted CMB
angular power spectrum for the case with $n=4$ normalizing the
Sachs-Wolfe tail.  Here, we use the relation (\ref{eq:kappam}) with
$\cot n\theta_{\rm init}=0$ which makes the discrepancy smallest.  As
one can see, even in the case with $\cot n\theta_{\rm init}=0$, the
CMB angular power spectrum becomes extremely inconsistent with the
WMAP results.  Indeed, the total angular power spectrum is
characterized by the following $\kappa$ parameters:
\begin{eqnarray}
    \kappa_{\rm corr} \simeq -\frac{9}{4} 
    \frac{(n-2)\Omega_b}{\Omega_m},~~~
    \kappa_{\rm iso} \simeq \frac{9}{4} \frac{\Omega_b}{\Omega_m}
    \sqrt{(n-2)^2 + n^2 \cot^2 \left(n \theta_{\rm init} \right)},
\end{eqnarray}
and, using $n\geq 3$ and adopting our canonical values of $\Omega_b$
and $\Omega_m$, we obtain $\kappa_{\rm corr}\lesssim -0.39$ and
$|\kappa_{\rm iso}| > 0.39$, which is inconsistent with the
observational results.  (See Fig.~\ref{fig:kappa-const}).

Therefore, the present case is excluded by the WMAP result. As we have
shown, this is due to too large entropy fluctuation in the baryonic
sector~\cite{EKM}. We should emphasize that the result obtained here is quite
generic, independently of the detailed dynamics of the Affleck-Dine
mechanism as far as the Affleck-Dine field is an $F$- and $D$-flat
direction.  Thus, we conclude that the Affleck-Dine field cannot play
the role of the curvaton.  In the next subsection, we will see that
essentially the same conclusion is obtained for the case with the
$Q$-ball formation.

\subsection{Case with $Q$-ball formation}
\label{sec:wQ}

If the potential of the Affleck-Dine field is flatter than the
parabolic one, it is inevitable that the $Q$-balls are formed.  This
is due to the fact that, if $K<0$, instability band arises in the
momentum space, which is given by\footnote
{In fact, the numerical factor for the instability band depends on the
value of $\epsilon$, but the order of magnitude does not change.}
\begin{eqnarray}
   0 < k_{\rm phys}^2 < 2 m_\phi^2 |K|,
   \label{k(instability)}
\end{eqnarray}
where $k_{\rm phys}$ is the physical momentum.  In this case, once the
Affleck-Dine field starts to oscillate, fluctuation with the momentum
within the instability band grows rapidly.

Typical initial charge of the $Q$-ball generated in this process
depends on the value of $\epsilon$.  If $\epsilon$ is close to 1,
charge of the typical $Q$-ball is given by the total charge within the
horizon at the time of the $Q$-ball formation.  As a result, typical
initial charge $|Q|_{\rm init}$ is proportional to $\epsilon$.  In
this case, formation of the $Q$-ball with opposite sign of charge
(so-called anti-$Q$-ball) is ineffective.  If $\epsilon$ becomes much
smaller than 1, however, $|Q|_{\rm init}$ becomes independent of
$\epsilon$.  In this case, almost same number of the $Q$-ball and
anti-$Q$-ball are formed with the relation
$\epsilon=(n_Q-n_{\bar{Q}})/(n_Q+n_{\bar{Q}})$, where $n_Q$ and
$n_{\bar{Q}}$ are the number densities of the $Q$-ball and
anti-$Q$-ball, respectively.  Indeed, numerical lattice simulations
have shown that
\begin{eqnarray}
   |Q|_{\rm init} \sim \bar{\beta}_n 
   \left( \frac{\phi_{\rm init}}{m_\phi} \right)^2 \times
   \left\{ \begin{array}{ll}
   \epsilon 
   ~~~&~~~ {\rm for}\ \epsilon \gtrsim \epsilon_{\rm c}\\
   \epsilon_{\rm c}
    ~~~&~~~ {\rm for}\ \epsilon \lesssim \epsilon_{\rm c}
   \end{array} \right. ,
   \label{Q-bar}
\end{eqnarray}
where $\bar{\beta}_n \sim 6 \times 10^{-3}$ and $\epsilon_{\rm c}\sim
0.01$ \cite{KasKaw}.\footnote
{In fact, the charges of produced $Q$-balls distribute with typical
value given by Eq.\ (\ref{Q-bar}). The precise estimation of the
distribution is difficult since the $Q$-balls are produced through
meta-stable state ``$I$-ball'' \cite{iball} and it needs very long
time simulations to find the distribution of the final stable
$Q$-balls.  However, the dynamics only depends on $\phi_{\rm init}$,
{\it i.e.}, which justifies the present analysis.}

In addition, the condition for the instability band
(\ref{k(instability)}) provides the size of the $Q$-ball as
\begin{eqnarray}
   R_Q \sim \frac{1}{|K|^{1/2} m_\phi}.
\end{eqnarray}
%%
%\rem{factor 2???}
As long as the initial value of the $Q$-ball charge is larger than
$O(10^{18})$, evaporation of the $Q$-ball is ineffective and the
universe is reheated by the decay of the $Q$-ball.  The decay process
of a single $Q$-ball is described by 
\begin{eqnarray}
   \dot{Q} \sim - \frac{m_\phi^3 {\cal A}}{192\pi^2},
\end{eqnarray}
where ${\cal A}=4\pi R_Q^2$ is the surface area of the $Q$-ball
\cite{NPB272-301}.  (This relation is valid when the charge of the
$Q$-ball is much larger than 1.)  Thus, in our case, $\dot{Q}$ is
(almost) a constant of time and hence the lifetime of the $Q$-ball
with initial charge $Q_{\rm init}$ is given by $\Gamma_Q^{-1}\sim
|Q_{\rm init}/\dot{Q}|$.  Using Eq.\ (\ref{Q-bar}) with
$\epsilon\lesssim \epsilon_c$, the decay temperature of the $Q$-ball,
{\it i.e.}, $T_{\rm RD2}$, is estimated as
\begin{eqnarray}
    T_{\rm RD2} \sim 1\ {\rm MeV} \times
    \left( \frac{|K|}{0.03} \right)^{-1/2}
    \left(\frac{m_\phi}{1 {\rm TeV}} \right)^{3/2}
    \left( \frac{\phi_{\rm init}}{M_*} \right)^{-1}.
    \label{T_RD2(Qball)}
\end{eqnarray}
One of important consequences of the $Q$-ball formation is that the RD2
epoch may be realized at much lower temperature compared to the case
without $Q$-ball formation; for the $Q$-ball charge $Q_{\rm init}\sim
10^{21} - 10^{27}$, the RD2 epoch is realized at the temperature
$T_{\rm RD2}\sim 1\ {\rm MeV}-1\ {\rm GeV}$, which is lower than the
freeze-out temperature of the lightest neutralino (which is assumed to
be the LSP). The other important consequence is that the decay rate of
$Q$-balls, $\Gamma_Q$, can be another source of the density fluctuations
\cite{dGamma2Psi}.
%Dvali:2003em,Kofman:2003nx,Dvali:2003ar,Matarrese:2003tk}.
Note that $\Gamma_Q$ depends on $\phi_{\rm init}$,
and that the decay process is different from the usual exponential decay.
Hence, if $\phi_{\rm init}$ fluctuates spatially, $\Gamma_Q$ has spatial
variation, leading to additional density fluctuations.

Now we discuss the density fluctuation in the case with $Q$-ball
formation.  First, we consider the metric perturbation.  Taking
account of the fact that fluctuation of the decay rate generates
adiabatic density perturbation \cite{dGamma2Psi}, the metric
perturbation has the following form:
%Now we discuss the density fluctuations in the case with $Q$-ball
%formation.  First we consider the metric perturbation, which can be
%expressed as
%%
\begin{eqnarray}
    \left[ \Psi^{(\delta\phi_r)} \right]_{\rm RD2} \simeq 
    -\left(\frac{4}{9}+C\right) \frac{\delta\phi_r}{\phi_{\rm  init}},~~~
    \left[ \Psi^{(\delta\phi_\theta)} \right]_{\rm RD2} \simeq 0,
\end{eqnarray}
where the second term in  $\Psi^{(\delta\phi_r)}$ represents 
the contribution due to the varying decay rate, and the typical
value of $C$ is estimated to be order of unity.
For our purpose, it is not necessary to know the precise 
value of $C$, but we crudely set $C=1$. As long as $C \sim O(1)$, our
conclusion does not change. Note that its sign is determined by
the fact that the lifetime of $Q$-balls becomes longer for larger $\phi_{\rm init}$,
leading to more negative gravitational potential.
Next we consider the isocurvature fluctuation in the baryonic and CDM sector.
Since the density fluctuation due to the varying decay rate does not
change entropy perturbations, we take $\Gamma_Q$ constant in the following.
While the baryonic isocurvature fluctuation is
%Even in this case, the metric perturbation and the
%isocurvature fluctuation in the baryonic sector are
calculated as in the previous subsection,  the property of $\delta_c$ can be
different from the case without $Q$-ball formation.  

To discuss the isocurvature fluctuation in the CDM sector, 
it is important to understand how the relic density of the
lightest neutralino ({\it i.e.}, the CDM density) is determined.
If $T_{\rm RD2}$ is lower than the freeze-out temperature of the LSP,
the relic abundance of the LSP is non-thermally determined.  The relic
abundance depends on whether the annihilation rate of the LSP at the
decay is larger or smaller than the expansion rate of the universe.
If the annihilation rate is smaller than the expansion rate, (almost)
all the LSPs produced by the decay of the $Q$-ball survive.  In this
case, no entropy fluctuation is expected between the LSP ({\it i.e.},
the CDM) and the photon produced by the $Q$-ball decay.  If all the
LSPs produced by the $Q$-ball survives, however, the relic density of
the LSP is expected to become much larger than the critical density of
the universe.  This is because the $Q$-ball consists of the
condensation of supersymmetric particles ({\it i.e.}, the scalar
quarks and/or scalar leptons) and hence the situation corresponds to
the case discussed in Section \ref{sec:moduli} with
$\bar{N}_{\tilde{\chi}}\sim 0.1-1$.  Indeed, from Eq.\
(\ref{eq:rhocs}), we can estimate the LSP density normalized by the
entropy density of the universe:
\begin{eqnarray}
    \frac{\rho_c}{s} &\simeq & 10^{-5} {\rm GeV} 
    \left(\frac{\bar{N}_{\tilde{\chi}}}{0.1}\right)
    \left(\frac{T_{\rm RD2}}{1{\rm MeV}}\right)
    \left(\frac{m_{\tilde{\chi}}}{100{\rm GeV}}\right)
    \left(\frac{m_{\phi}}{1{\rm TeV}}\right)^{-1},
\end{eqnarray}
which is much larger than the observed value $\rho_{\rm crit}/s\simeq
3.6 \times 10^{-9} h^2 {\rm GeV}$, with $\rho_{\rm crit}$ being the
critical density of the universe.

Thus, in the following we discuss the case where the annihilation rate
of the LSP is larger than the expansion rate of the universe.  In this
case, the pair annihilation proceeds until the annihilation rate
becomes comparable to the expansion rate and hence
$[n_{\tilde{\chi}}]_{\rm H\sim\Gamma_Q}\sim\Gamma_Q/\langle v_{\rm
rel}\sigma\rangle$.  In the RD2 epoch or later, ratio of
$n_{\tilde{\chi}}$ to $n_\gamma$ is conserved and hence we obtain
\begin{eqnarray}
   \frac{n_{\tilde{\chi}}}{n_\gamma} \sim 
   \frac{1}{T_{\rm RD2}M_* \langle v_{\rm rel}\sigma\rangle},
\end{eqnarray}
where we have used the relation $\Gamma_Q\sim T_{\rm RD2}^2/M_*$.  We
assume that $\Gamma_Q$ and $\langle v_{\rm rel}\sigma\rangle$ satisfy
the relation such that the above ratio becomes consistent with the
currently observed dark matter density~\cite{Fujii-Hamaguchi}.  

%\rem{More later.  Reference@to Fujii-Hamaguchi.}

As shown in Eq.\ (\ref{T_RD2(Qball)}), $T_{\rm RD2}$ depends on
$\phi_{\rm init}$ and hence the ratio $n_{\tilde{\chi}}/n_\gamma$
fluctuates if the Affleck-Dine field has primordial fluctuation.
Using $T_{\rm RD2}\propto\phi_{\rm init}^{-1}$, we obtain
$n_{\tilde{\chi}}/n_\gamma\propto\phi_{\rm init}$.  Then,
\begin{eqnarray}
   S_{c\gamma}^{(\delta\phi_r)} = 
   \frac{\delta\phi_r}{\phi_{\rm init}},~~~
   S_{c\gamma}^{(\delta\phi_\theta)} = 0,
\end{eqnarray}
where we have used the relation
$S_{c\gamma}=\delta(n_c/n_\gamma)/(n_c/n_\gamma)$.  In this case, the
fluctuation in the radial direction, $\delta\phi_r$, also generates
the isocurvature fluctuation in the CDM sector.  Notice that the
fluctuation in the phase direction $\delta\phi_\theta$ does not affect
the isocurvature fluctuation in the CDM since $\epsilon\ll 1$.

In the case of the $Q$-ball formation, significant amount of the
isocurvature fluctuations are generated associated with $\delta\phi_r$ and
$\delta\phi_\theta$.  Adding the isocurvature fluctuations in the baryon
and the CDM, we obtain
\begin{eqnarray}
    \kappa_m^{(\delta\phi_r)} \simeq 
    -\left(\frac{4}{9}+C\right)^{-1}
    \frac{(n-2)\Omega_b + \Omega_c}{\Omega_m},~~~
    \kappa_m^{(\delta\phi_\theta)} \simeq \infty.
    \label{eq:kappam-wQ}
\end{eqnarray}
Importantly, $\kappa_m^{(\delta\phi_r)}$ is again negative and
$\kappa_m^{(\delta\phi_\theta)}\gg 1$.
Thus, $\delta\phi_r$ results in the density fluctuations with
correlated mixture of the adiabatic and isocurvature fluctuations
while $\delta\phi_\theta$ becomes (almost) purely isocurvature
fluctuations. The $\kappa$ parameters, which characterize the
total CMB power spectrum, are thus given by
\begin{eqnarray}
	\label{eq:kappa_wQ}
    \kappa_{\rm corr} &\simeq& -\left(\frac{4}{9}+C\right)^{-1} 
    \frac{(n-2)\Omega_b+\Omega_c}{\Omega_m},\non\\
        \kappa_{\rm iso} &\simeq&\left(\frac{4}{9}+C\right)^{-1}  \Omega_m^{-1}
    \sqrt{((n-2)\Omega_b+\Omega_c)^2 +( n \cot \left(n \theta_{\rm init} \right) \Omega_b)^2}.
\end{eqnarray}
%%

%, and the situation becomes worse
%than the previous case due to the entropy fluctuation in the CDM
%sector.  
For the total CMB angular power spectrum, if one normalizes
the Sachs-Wolfe tail, acoustic peaks are extremely suppressed compared
to the purely adiabatic case.  In Fig.\ \ref{fig:ad_cmb}, the CMB
angular power spectrum with the relation Eq.~(\ref{eq:kappa_wQ}) is also
plotted in the case of $\cot n\theta_{\rm init}=0$ and $C=0$ and $1$.  
In this case, again, the shape of the total CMB angular power spectrum 
becomes inconsistent with the observations.

To summarize, the Affleck-Dine field cannot be the curvaton in both of
the cases with and without $Q$-ball formation because the entropy
density fluctuation in baryonic and/or CDM sector is too large.  In
other words, if Affleck-Dine field (or $Q$-ball) dominates the energy
density of the universe, its primordial fluctuation must be
suppressed,\footnote
{In fact, the primordial fluctuation of the Affleck-Dine field is
suppressed if there are Hubble-induced terms in the potential and/or
the initial amplitude is large enough.}
and the dominant density fluctuation must be generated by another
source, like inflaton. 

\section{$F$- and $D$-Flat Direction as Curvaton}
\label{sec:flatdir}
\setcounter{equation}{0}

In this section, we consider the case where an $F$- and $D$-flat
direction plays the role of curvaton which is not related to the
mechanism of baryogenesis.  Even in this case, $Q$-ball may or may not
be produced depending on the shape of the curvaton potential; in
particular, if the curvaton potential is flatter than parabolic (which
may be due to the renormalization effects), $Q$-ball can be produced
once the curvaton starts to oscillate.  As can be expected from the
discussion given in the previous section, properties of the
cosmological density fluctuations strongly depend on how the curvaton
field evolves.

First, let us consider the simplest possibility, namely, the case
without the $Q$-ball formation.  In this case, $T_{\rm RD2}$ is
expected to be higher than the freeze out temperature of the lightest
supersymmetric particle and we assume that the CDM is {\sl thermally}
produced lightest neutralino.  If $T_{\rm RD2}$ is high enough,
baryogenesis may be possible with the decay products of $\hat{\phi}$.
For example, the Fukugita-Yanagida mechanism \cite{PLB174-45} using
the decay of thermally produced right-handed neutrinos may work if
$T_{\rm RD2}\gtrsim 10^{9-10}~{\rm GeV}$~\cite{NPB643-367}.  Another
possibility may be electroweak baryogenesis \cite{PLB155-36} if
$T_{\rm RD2}\gtrsim 100\ {\rm GeV}$.

In this simplest case, no isocurvature fluctuation between the
non-relativistic components and the radiation is generated by
$\delta\phi_r$ nor $\delta\phi_\theta$ ({\it i.e.}, $\kappa_m=0$).
Thus, in the RD2 epoch, the only source of the cosmic density
fluctuations is the metric perturbation given in Eq.\ 
(\ref{Psi(RD2)}).  Since no isocurvature fluctuation is generated,
$C_l^{(\delta\phi_r)}$ and $C_l^{(\delta\phi_\theta)}$ are both from
purely adiabatic density fluctuations and hence
$C_l^{(\delta\phi_r)}\propto C_l^{(\delta\phi_\theta)}$.  Thus, even
though $\Psi^{(\delta\phi_r)}$ and $\Psi^{(\delta\phi_\theta)}$ are
free parameters, it does not affect the shape of the total CMB angular
power spectrum and $C_l$ is the same as the result with
purely-adiabatic scale-invariant primordial density fluctuations.
(Concrete models for the flat-direction curvaton is found in Ref.
\cite{FlatdirCurv}.)

As discussed in the previous section, if the scalar potential of the
flat direction field is flatter than the parabolic one, the $Q$-balls
are formed. If the initial charge of the $Q$-ball is small enough, it
decays (or evaporates) with a temperature above the freeze-out
temperature of the LSP. This case is reduced to the one discussed just
above, where the purely-adiabatic density fluctuation is obtained.  If
the charge of the $Q$-ball is large, however, it decays after the
thermal pair-annihilation of the LSPs are frozen out. Thus, one can
see that the entropy fluctuation is generated in the CDM sector:
\begin{eqnarray}
    S_{c\gamma}^{(\delta\phi_r)} = 
    \frac{\delta\phi_r}{\phi_{\rm init}},~~~
    S_{c\gamma}^{(\delta\phi_\theta)} = 0,
\end{eqnarray}
which leads to 
\begin{eqnarray}
    \kappa_m^{(\delta\phi_r)} \simeq 
    -\left(\frac{4}{9} + C\right)^{-1} \frac{\Omega_c}{\Omega_m},~~~
    \kappa_m^{(\delta\phi_\theta)} = 0,
\end{eqnarray}
where $C \sim O(1)$ represents the contribution due to the
varying decay constant. 
Notice that there is no baryonic entropy fluctuation, since the
curvaton has nothing to do with baryon asymmetry.\footnote{
However, if baryogenesis takes place before the $Q$-ball dominated
universe, large isocurvature fluctuation is induced in the baryon sector,
{\it i.e.}, $\kappa_b=\frac{9}{2}$. Here we assume that the baryon asymmetry
is somehow generated after the beginning of $Q$-ball-dominated
universe.
}  In this case,
$\left|\Psi^{(\delta\phi_\theta)}\right|\ll\left|\Psi^{(\delta\phi_r)}\right|$ and hence
$\delta\phi_\theta$ does not play any significant role in generating
the cosmic density fluctuations.  Thus, with the best-fit values of
$\Omega_c$ and $\Omega_m$ obtained from the WMAP experiment,
$\kappa_{\rm corr} \simeq -1.9, -0.57$ for $C=0,1$ respectively, 
which are clearly excluded by the
observation. (See Fig.~\ref{fig:kappa-const}).  Therefore, again, the
curvaton with $Q$-ball formation which decays after the freeze-out of
the LSP is rejected.

\section{Right-Handed Sneutrino as Curvaton}
\label{sec:snu}
\setcounter{equation}{0}

Recent neutrino-oscillation experiments strongly suggest tiny but
non-vanishing masses of the neutrinos.  Such small masses can be
naturally explained by introducing the right-handed neutrinos
\cite{seesaw}.

If the right-handed neutrino exists in supersymmetric models, its
superpartner may have non-vanishing primordial amplitude.  In
particular, such a case has been studied related to the possibility of
generating the lepton number (and baryon number) asymmetry of the
universe from the decay product of the right-handed sneutrino
\cite{snu-leptogen}.  In this case, the right-handed sneutrino can be an
another candidate for the curvaton \cite{hph0211019}.

The scenario with the right-handed sneutrino condensation is almost
the same as the previous cases except for the origin of the baryon
asymmetry of the universe.  During inflation, the right-handed
sneutrino is assumed to have a non-vanishing amplitude.  After
inflation, the right-handed sneutrino starts to oscillate when the
expansion rate of the universe becomes comparable to the sneutrino
mass $m_{\phi}$.  If the decay of the oscillating right-handed
sneutrino takes place much later than the inflaton decay, the
sneutrino dominates the universe.  Then, the sneutrino
decays when the expansion rate of the universe becomes comparable to
its decay rate.  At the time of the decay, lepton number asymmetry can be
generated provided that CP violation exists in the neutrino sector.
Once the lepton number asymmetry is generated, it can be converted to
the baryon number asymmetry due to the sphaleron process
\cite{PLB155-36}.

In this scenario, quantum fluctuations that the sneutrino acquires
during inflation accounts for the adiabatic density fluctuations and
hence the right-handed sneutrino plays the role of the curvaton.
Here, baryon asymmetry of the universe is generated from the decay
products of the right-handed sneutrino and hence no isocurvature
fluctuation is expected in the baryonic sector.  In addition, the
decay rate of the right-handed sneutrino is high enough so that the
LSPs (and other superpartners of the standard-model particles) are
thermally produced.  Then there is no isocurvature fluctuation in the
CDM sector if the LSP becomes the CDM.

If this is the case, the primordial density fluctuations become purely
adiabatic and hence the resultant CMB angular power spectrum becomes
well consistent with the results of the WMAP experiment.  Thus, we
conclude that the right-handed sneutrino is one of the good and
well-motivated candidates for the curvaton.

\section{Conclusions and Discussion}
\label{sec:conclusion}
\setcounter{equation}{0}

We have studied several candidates for the curvaton in the
supersymmetric framework.  Since scalar fields appearing in
supersymmetric models are inevitably complex, it is necessary to
consider cases with multi-curvaton fields.  In this case, primordial
fluctuations in the curvatons induce isocurvature (entropy)
fluctuations as well as the adiabatic ones.  Such isocurvature
fluctuations may affect the CMB angular power spectrum and the recent
WMAP experiment sets stringent constraints on those models.

One potential problem of the curvaton scenario in the supersymmetric
models is to suppress the baryonic isocurvature fluctuations.  As
shown in the present paper, isocurvature fluctuation in the baryonic
sector may arise in various situations.  In many cases, such a
baryonic isocurvature becomes too large to be consistent with the WMAP
result.

Besides baryonic isocurvature fluctuations, we have found that the CDM
isocurvature fluctuations are important for moduli or $Q$-ball
curvaton where the abundance of the dark matter ({\it i.e.}, the LSP)
is determined non-thermally and hence its fluctuations are
isocurvature with correlation to the curvaton fluctuations.

If the cosmological moduli fields play the role of the curvaton, it is
possible that the reheating temperature due to the decay of the moduli
fields is lower than the freeze-out temperature of the
lightest neutralino.  In this case, the CDM is non-thermally produced
if the lightest neutralino is the CDM, and the entropy fluctuation
between the two curvatons ({\it i.e.}, the real and imaginary parts of
the curvaton) in general becomes the isocurvature fluctuation in the
CDM sector.  This fact excludes some part of the parameter space but,
as we have seen, slight hierarchies between the initial amplitudes
and/or the decay rates of two curvaton fields are enough to make the
resultant CMB anisotropy consistent with the WMAP result.  Of course,
if the lightest neutralino is not the CDM, then we may evade the
constraint from the isocurvature fluctuation in the CDM sector.  An
example for such CDM candidates is axion.

If the Affleck-Dine field becomes the curvaton, on the contrary, too
large baryonic isocurvature fluctuation is inevitably induced.  As we
have seen, the acoustic peaks are always too much suppressed to be
consistent with the observations if the Affleck-Dine field plays the
role of the curvaton.  Thus, the Affleck-Dine field
cannot play the role of the curvaton.  It is also notable that, in the
case with $Q$-ball formation, extra isocurvature fluctuation may arise
if the reheating temperature due to the $Q$-ball decay is lower than
the freeze-out temperature of the lightest neutralino (which is
assumed to be the CDM).

For the case where the $F$- and $D$-flat direction (without baryon
number) becomes the curvaton, isocurvature fluctuation in the CDM
sector may also arise if the $Q$-ball is formed.  In this case, the
resultant CMB angular power spectrum is too much affected by the
isocurvature fluctuation.  Without the $Q$-ball formation, however,
the $F$- and $D$-flat direction (but not the Affleck-Dine field) can
be a candidate of the curvaton.

Another good and well-motivated candidate of the curvaton is the
right-handed sneutrino.  If the right-handed sneutrino acquires the
primordial quantum fluctuations during inflation, it becomes purely
adiabatic density fluctuations and the resultant CMB angular power
spectrum can be well consistent with the WMAP result.

In summary, for a viable scenario of the curvaton,
 the isocurvature contribution should be so small that the CMB
angular power spectrum becomes consistent with the WMAP result.  In
the supersymmetric case, this fact provides stringent constraints on
the curvaton scenario.  Thus, it is necessary to look for curvaton
candidates which do not generate the isocurvature fluctuations; some
of them are already found in the present study.

{\sl Acknowledgments:} 
We acknowledge the use of CMBFAST \cite{cmbfast} package for our
numerical calculations.  The work of T.M. is supported by the
Grant-in-Aid for Scientific Research from the Ministry of Education,
Science, Sports, and Culture of Japan, No.\ 15540247.  K.H. and
F.T. thank the Japan Society for the Promotion of Science for
financial support.

%%%%%%%%%%%%%%%%%%%%%%%%%%%%%%%%%%%%%%%    
%%%%%%%%%%%%%%%%%%
%%%%%%%%%%%%%%%%%%%%%%%%%%%%%%%%%%%%%%%    

\end{document}